%% file: main.tex
\documentclass{article}
\PassOptionsToPackage{numbers}{natbib}
\usepackage[main,final]{neurips_2026}

\input{math_commands.tex}

\usepackage{hyperref}
\usepackage{url}
\usepackage{graphicx}
\usepackage{soul}
\usepackage{xcolor}
\usepackage{xspace}
\usepackage{tabularray}
\UseTblrLibrary{booktabs}

\newcommand{\projectname}{\textsc{FireFly}\xspace}

\title{\raisebox{-0.2\height}{\includegraphics[width=1.5em]{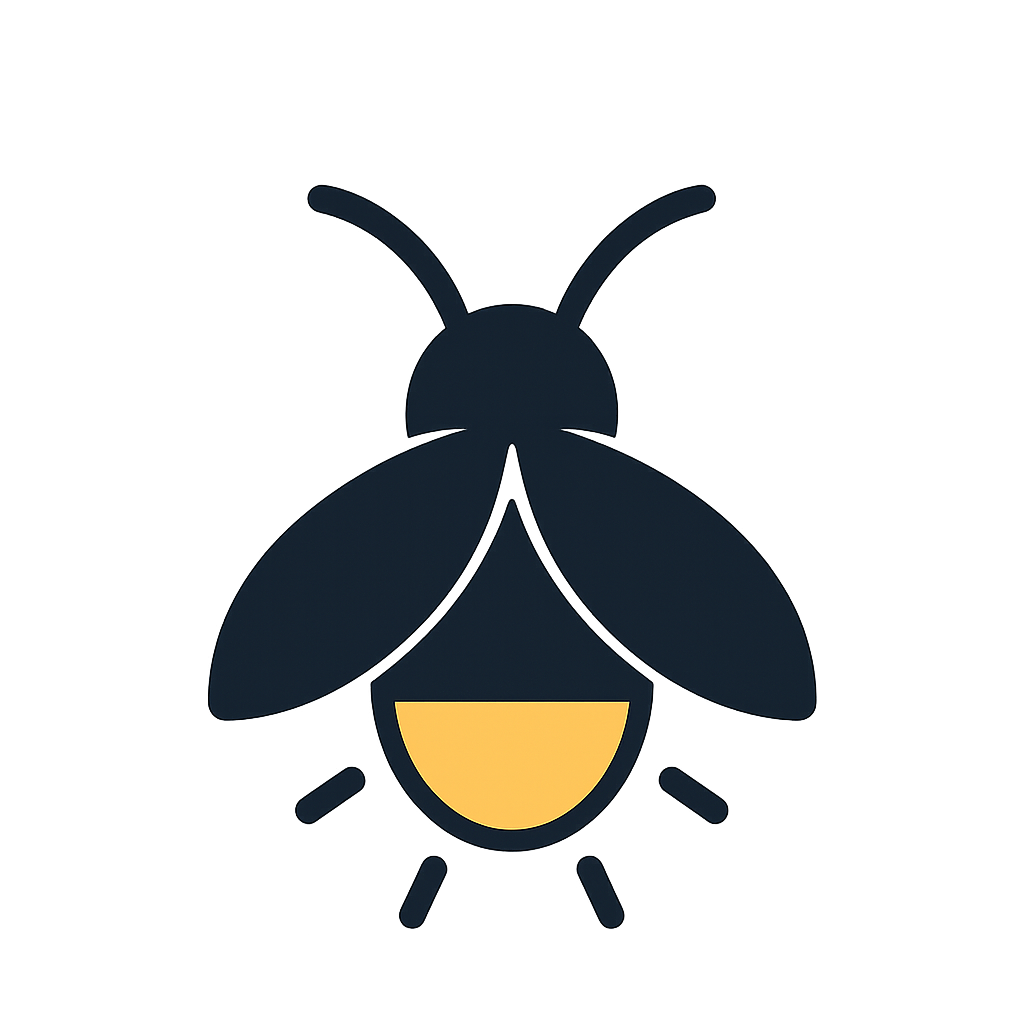}}~Firefly: Illuminating Large-Scale Verified Tool-Call Data Generation from Real APIs}

\author{%
  Yuxuan Lu$^{1}$\thanks{\texttt{lu.yuxuan@northeastern.edu}.},\;
  Ziyi Wang$^{1}$,\;
  Yingzhou Lu$^{2}$,\;
  Yisi Sang$^{3}$,\;
  Jiri Gesi$^{3}$,\;
  Xianfeng Tang$^{3}$,\; \\ \bf
  Yimeng Zhang$^{3}$,\;
  Zhenwei Dai$^{3}$,\;
  Hui Liu$^{3}$,\;
  Hanqing Lu$^{3}$,\;
  Chen Luo$^{3}$,\;
  Qi He$^{3}$,\; \\ \bf
  Benoit Dumoulin$^{3}$,\;
  Jing Huang$^{3}$,\;
  Dakuo Wang$^{1}$\thanks{Corresponding author: \texttt{d.wang@northeastern.edu}}\\[6pt]
  \normalfont $^{1}$Northeastern University,\quad $^{2}$Stanford University,\quad $^{3}$Independent Researcher
}

\begin{document}

\maketitle

\begin{abstract}
Training tool-calling agents requires large-scale trajectory data with verifiable labels, yet existing approaches either synthesize environments that diverge from real API behavior or generate tasks without ground-truth outcomes for verification.
We present \projectname, a pipeline for generating verified tool-call data from real-world MCP servers. Our key insight is to invert the standard synthesis pipeline: rather than generating tasks and hoping they are solvable, we first let a strong LLM explore real APIs along graph-guided DAG structures, then synthesize tasks backward from observed outcomes, guaranteeing label correctness by construction.
To handle the scale of real-world tool spaces (${\sim}$1,000 tools), we build a pairwise tool graph and sample sub-DAGs to focus exploration on semantically coherent workflows.
To address environment drift in live APIs, we construct a retrieval-augmented simulator that caches all exploration results and replays them during training and evaluation, enabling reproducible RL without live server access. Applying this pipeline yields 5,144 verified tasks spanning 240 servers and 993 tools. A 4B-parameter model trained with GRPO on \projectname~matches Claude Sonnet 4.6 on our held-out test set and shows improvements on multiple tool-calling benchmarks including Tau2-Bench, MCPMark, and MCP-Atlas. Code is available at \url{https://github.com/leoleoasd/firefly-neurips-2026/}, and data and model are available at \url{http://firefly.hailab.io/}.
\end{abstract}

\begin{figure}[h]
    \centering
    \includegraphics[width=\textwidth]{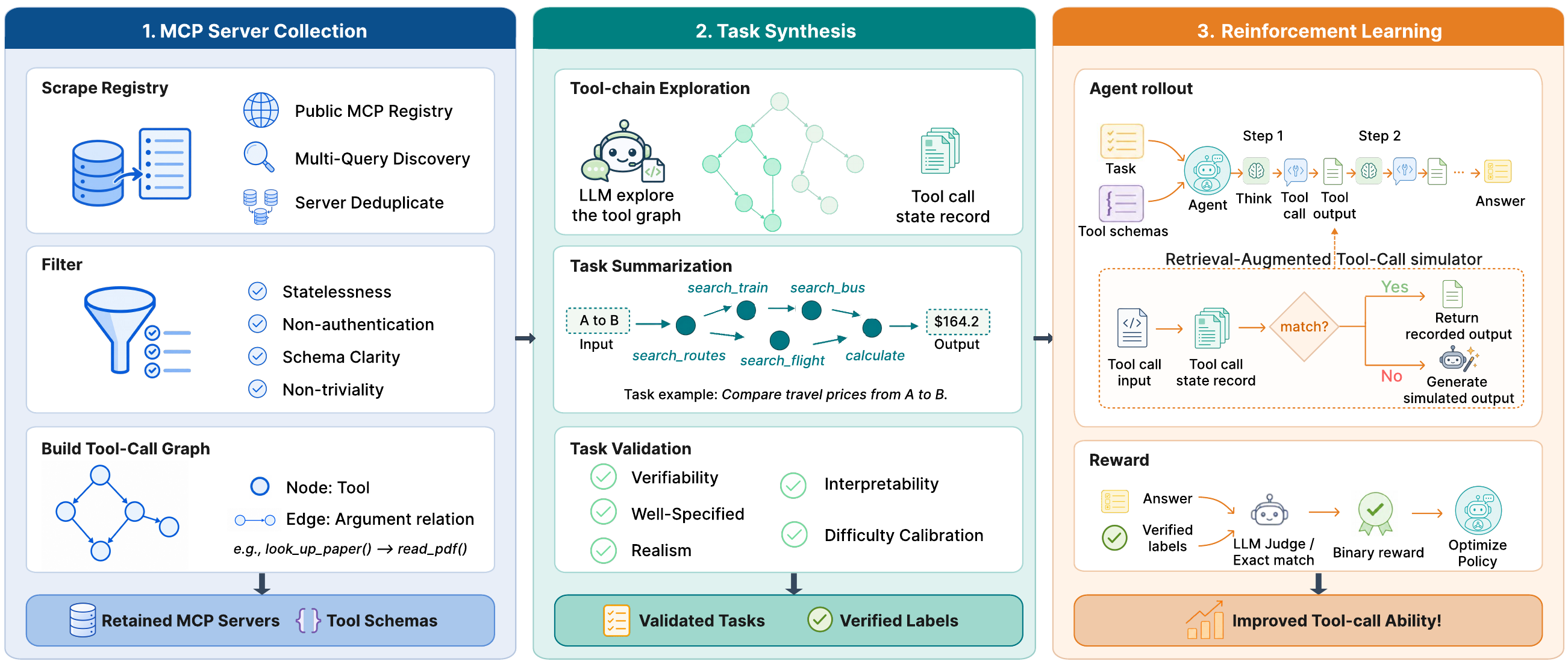}
\caption{
Overview of \projectname.
The pipeline first collects real-world MCP servers, filters them for reproducible and benchmarkable tool use, and constructs a tool-call graph from tool schemas.
It then explores valid tool chains and summarizes observed tool-call states into natural-language tasks with verified labels, followed by validation for task quality and reliability.
Finally, we use the validated tasks and verified labels for reinforcement learning: to avoid unstable live MCP servers during rollout, we build a retrieval-augmented tool-call simulator that replays or simulates tool outputs offline, enabling stable RL training with binary rewards and improving tool-call ability.
}
    \label{fig:pipeline}
\end{figure}

\section{Introduction}

Tool-calling agents interact with external APIs to complete user intents, and they have become a central paradigm for building AI systems that can act in real environments~\citep{schickToolformerLanguageModels2023,yaotaubenchBenchmarkToolAgentUser2024,barrestau2BenchEvaluatingConversational2025,wangTrajectory2TaskTrainingRobust2026}.
However, training such agents requires many tasks with correct intermediate tool calls and verifiable final outcomes.
Even data-efficient reinforcement learning recipes consume thousands of such tasks, for example 2.4k training samples for a search agent~\citep{jiangS3YouDont2025}, whereas recent agent benchmarks each offer only tens to hundreds, such as 50 to 115 tasks per domain in Tau2-Bench~\citep{barrestau2BenchEvaluatingConversational2025}, 127 tasks in MCPMark~\citep{wuMCPMarkBenchmarkStressTesting2025}, and 500 public tasks in MCP-Atlas~\citep{bandiMCPAtlasLargeScaleBenchmark2026}.
Human annotation could close this gap, but it is expensive, domain-specific, and difficult to scale to the long tail of real APIs~\citep{yaotaubenchBenchmarkToolAgentUser2024,luLLMsNowFineTuning2026a, chenStorySparkQAExpertAnnotatedQA2024a, chenMultiAgentasJudgeAligningLLMAgentBased2026, chenHumansALMANACHuman2026}.
Synthetic pipelines scale more easily, although each existing design gives up either realism or verifiability.
When they rely on purely simulated environments, an LLM invents tool responses that often diverge from real API behavior~\citep{chenScalingAgentLearning2025}.
When they instead generate a user task first and then ask an LLM to produce a trajectory over real tools~\citep{tangToolAlpacaGeneralizedTool2023,prabhakarAPIGenMTAgenticPipeline2025a,xuTOUCANSynthesizing15M2025}, the trajectory may be infeasible and the final answer is often not tied to any observed state, so the task lacks a ground truth that could verify correctness~\citep{xuTOUCANSynthesizing15M2025}.
Moreover, even verified tasks over real tools are hard to train on, because live APIs drift over time, become unavailable, and rate-limit the repeated calls of RL rollouts.
As a result, constructing training tasks that are scalable, realistic, and verifiable, together with an environment that stays reproducible during RL, remains an open problem.

To address the verifiability problem, \projectname inverts the forward pipeline.
Specifically, we first explore real tools, record the states reached through actual execution, and only then synthesize tasks backward from these observed outcomes.
In this back-chaining process, every label is correct by construction because it records a state that real tools actually returned.
Concurrent work applies the same trajectory-to-task idea in small or fixed tool spaces, where the set of possible tool chains can be enumerated or manually constrained~\citep{wangTrajectory2TaskTrainingRobust2026}, so the open question is whether back-chaining works at the scale of real-world MCP ecosystems.

Scaling back-chaining to real MCP servers introduces a challenge that small tool spaces avoid, because the number of possible tool combinations grows quickly with the number of tools.
In our setting, hundreds of servers expose nearly one thousand tools across heterogeneous domains, so naive exploration wastes most calls on semantically incoherent tool combinations.
To make exploration scalable, \projectname constructs a directed tool graph over all tools, where edges represent pairwise input-output compatibility judged from tool schemas and descriptions.
We then sample sub-DAGs from this graph and let a strong LLM explore within these graph-guided neighborhoods.
The graph thus acts as a weak semantic prior that leaves the final task open while steering exploration toward coherent multi-tool chains that are likely to produce useful, compositional trajectories.

A second challenge is the environment drift of live APIs, which makes real tools unstable training environments.
Many MCP tools depend on external data sources such as search indices, weather, finance data, DNS records, package metadata, or live web services, so their outputs may change over time, servers may become unavailable, and repeated calls during RL rollout may be rate-limited or non-reproducible.
\projectname addresses this drift by caching every tool call observed during exploration, including the tool name, input parameters, and returned output.
We then build a replay-based, retrieval-augmented tool-call simulator that replaces live servers during training and evaluation.
Unlike purely simulated environments, this simulator grounds every response in real API outputs.
For exact matches it replays the cached response, and for near matches an LLM either returns the matching cached response or generates one conditioned on the closest cached outputs, whereas for calls that differ semantically from every cached call it returns an error.
The error case is intentionally conservative, because each ground-truth task was synthesized from a known trajectory, so calls far outside the cached trajectories are unlikely to be necessary for the correct solution and should not be rewarded (Sec.~\ref{sec:limitations} quantifies the resulting truncation risk).
As a result, \projectname keeps tool responses grounded in real API execution while providing the reproducibility and scalability required for simulator-based online RL.

Putting these ideas together, \projectname follows a three-stage pipeline.
First, it collects and filters real MCP servers, constructs a pairwise tool-compatibility graph, and performs graph-guided exploration to obtain executable tool-call DAGs.
Second, it synthesizes natural-language tasks by back-chaining from observed tool outputs and validates the resulting task-trajectory pairs for verifiability, specificity, realism, and difficulty.
Third, it builds a retrieval-augmented simulator from the cached exploration traces and uses it as the environment for simulator-based online RL training and for evaluation.
Applying this pipeline yields 5,144 verified tasks and 9,749 trajectories spanning 240 servers and 993 tools.
Training a 4B-parameter model with GRPO~\citep{shaoDeepSeekMathPushingLimits2024} and dynamic filtering~\citep{yuDAPOOpenSourceLLM2025} on \projectname improves pass@1 on the \projectname held-out test set from 28.1\% to 41.5\%, matching Claude Sonnet 4.6 under the same evaluation protocol.
The resulting model also transfers to public tool-calling benchmarks, improving all evaluated Tau2-Bench domains~\citep{yaotaubenchBenchmarkToolAgentUser2024,barrestau2BenchEvaluatingConversational2025}, MCP-Atlas~\citep{bandiMCPAtlasLargeScaleBenchmark2026}, and MCPMark~\citep{wuMCPMarkBenchmarkStressTesting2025} tasks.

Our contributions are threefold:
\begin{itemize}
    \item We introduce \projectname, a large-scale verified tool-calling dataset grounded in real MCP servers, together with a retrieval-augmented simulator that enables reproducible training and evaluation without live servers\footnote{\noindent Code is available at \url{https://github.com/leoleoasd/firefly-neurips-2026/}, and data and model are available at \url{http://firefly.hailab.io/}.}.
    \item We propose a scalable data-generation framework that combines graph-guided exploration with back-chaining task synthesis, making it possible to construct verified trajectories across hundreds of heterogeneous real-world API servers.
    \item We empirically show that RL training on \projectname substantially improves a 4B model on in-distribution verified MCP tasks and transfers to out-of-distribution tool-calling benchmarks such as Tau2-Bench~\citep{yaotaubenchBenchmarkToolAgentUser2024,barrestau2BenchEvaluatingConversational2025}, MCP-Atlas, and MCPMark.
\end{itemize}

\section{Related Work}

\subsection{Synthetic Data Generation}
\label{sub_sec:synthetic_data_gen}
The paradigm of LLM training with synthetic data has evolved from general instruction tuning to complex agentic tasks. Early works demonstrated the effectiveness of distilling strong models to generate diverse instruction-following data, significantly improving the capabilities of smaller models~\citep{xuSurveyKnowledgeDistillation2024, wangSelfConsistencyImprovesChain2023, xuWizardLMEmpoweringLarge2025}. 
Building on this success, recent research has pivoted towards synthesizing data specifically for reasoning and planning, employing techniques such as Chain-of-Thought (CoT) expansion and process supervision to enhance model robustness~\citep{weiChainThoughtPromptingElicits2023, luoWizardMathEmpoweringMathematical2025}.
In the domain of tool-calling agents, synthetic data is particularly critical due to the lack of high-quality, real-world API interaction logs. 
Pioneering works like ToolFormer~\citep{schickToolformerLanguageModels2023} and ToolAlpaca~\citep{tangToolAlpacaGeneralizedTool2023} introduced self-instruct mechanisms to generate API calls embedded within text. 
More recently, studies have focused on scaling up these datasets to cover broader domains. 
For instance, APIGen-MT~\citep{prabhakarAPIGenMTAgenticPipeline2025a} and similar frameworks~\citep{wangMCPFlowFacilitatingLLM2026, qi2025bear} utilize iterative verification processes to filter out invalid tool calls, aiming to improve the executable rate of generated actions. 
Concurrently, datasets like Toucan~\citep{xuTOUCANSynthesizing15M2025} have attempted to model complex tool dependencies by simulating multi-turn interactions.
Despite these advancements, existing synthetic frameworks face significant challenges regarding \textit{verifiability} and \textit{realism}. 
Most prior approaches rely on `forward generation', where the LLM is prompted to hallucinate both the user query and the subsequent tool trajectory. 
This often leads to plausible-looking but functionally incorrect calls, or `simulated' environments that fail to reflect the stochasticity and constraints of real-world servers. 
Furthermore, while execution-based filtering has been successfully applied in code generation via unit tests~\citep{chenEvaluatingLargeLanguage2021, luMachineLearningSynthetic2025, roziereCodeLlamaOpen2023}, applying rigorous verification to general tool-calling remains an open problem. 
Unlike coding tasks with deterministic outputs, tool interactions often lack ground-truth states for verification. 
\projectname~addresses these limitations by inverting the synthesis process: instead of hallucinating queries, we employ a \textit{back-chaining} strategy starting from valid, executed states on real Model Context Protocol (MCP) servers, ensuring that every generated task is grounded in a verifiable, executable trajectory.

\subsection{Tool-Calling Benchmarks and Datasets}
The transition of Large Language Models (LLMs) from static reasoning to interactive agentic behavior~\citep{wangSENTINELFailureDrivenReinforcement2026,chenRetrySwitchAbstain2026,luWEBSERVBrowserServerEnvironment2025a,chenMultiAgentasJudgeAligningLLMAgentBased2026,wangCustomerR1PersonalizedSimulation2025a,zhangShopr1RewardingLlms2026,chenHumansALMANACHuman2026,chenCollabSimCSCWGroundedMethodology2026} necessitates specialized benchmarks and datasets that accurately reflect real-world tool-calling scenarios.
Existing datasets for training and evaluating tool-calling agents are primarily categorized into two approaches.

First, \textit{human-curated datasets} are meticulously built by human annotators, exemplified by API-Bank~\citep{liAPIBankComprehensiveBenchmark2023}, which focuses on tool-calling planning, and the more recent Tau-Bench series~\citep{yaotaubenchBenchmarkToolAgentUser2024,barrestau2BenchEvaluatingConversational2025}, which aims for comprehensive coverage of diverse real-world APIs.
While human annotation guarantees the semantic validity and naturalness of the user intent, this approach is inherently resource-intensive, severely limiting scalability and making it difficult to maintain relevance across rapidly evolving API schemas.

Second, \textit{automatically generated datasets} have emerged to overcome these scale limitations, leveraging LLMs to synthesize tool-use instances.
Early efforts, such as ToolAlpaca~\citep{tangToolAlpacaGeneralizedTool2023}, focused on generating tool-augmented instruction data, while advanced frameworks such as APIGen-MT~\citep{prabhakarAPIGenMTAgenticPipeline2025a} and Toucan~\citep{xuTOUCANSynthesizing15M2025} use automated processes to synthesize tasks and expected trajectories.
However, these large-scale synthetic methods suffer from two critical shortcomings: the \textit{forward generation} paradigm leads to hallucinated tool calls and execution failures, requiring extensive filtering (as discussed in Sec.~\ref{sub_sec:synthetic_data_gen}); and many datasets are confined to simplified, \textit{simulated environments}~\citep{prabhakarAPIGenMTAgenticPipeline2025a}, failing to capture the complexity, constraints, and dynamic states of real-world MCP servers.

In summary, existing tool-calling datasets present a trade-off where no single benchmark simultaneously achieves verifiability, realism, and scalability.
\projectname closes this gap by grounding synthesis in real-world MCP servers and inverting the generation order to guarantee correctness.

\section{Method}
\label{sec:method}

Our pipeline constructs verifiable tool-use data in three phases: server collection (Sec.~\ref{sec:server_collection}), task synthesis (Sec.~\ref{sec:task_synthesis}), and trajectory generation during evaluation and rollout (Sec.~\ref{sec:trajectory_generation}).

\subsection{Server Collection}
\label{sec:server_collection}

\subsubsection{Scraping}
\label{sec:scraping}

We collect MCP servers from the Smithery registry\footnote{https://smithery.ai/} using broad listing queries, prefix enumeration, and domain-specific keyword searches. After deduplication and metadata fetching (tool definitions, input schemas, connection configurations), we obtain approximately 3,000 unique servers, of which roughly 900 expose at least one tool.

\subsubsection{Filtering}
\label{sec:filtering}

Many scraped servers are unsuitable for verified data construction. We screen each server against four criteria using an LLM evaluator:
\textbf{Statelessness} (no session memory or cross-call state),
\textbf{No user authentication} (no OAuth, credentials, or user-owned keys),
\textbf{Schema clarity} (explicit JSON typing and parameter descriptions), and
\textbf{Non-triviality} (no echo or diagnostic-only tools).
A server must meet all four. We spot-check a subset for alignment. 240 servers with 993 tools remain.

\subsubsection{Tool Graph}
\label{sec:tool_graph}

The tool graph is the mechanism that makes multi-step tasks possible in \projectname.
In an early pilot without the graph, the exploring LLM sampled tools at random across the full tool space, so successive calls were semantically unconnected and the output of one call was never usable as the input of the next.
As a result, essentially all tasks synthesized in that pilot collapsed to single-step lookups.
To obtain tool chains in which earlier outputs feed later inputs, we build a directed tool graph $G = (\mathcal{T}, \mathcal{E})$ over all retained tools across all servers. Each node represents a tool and each edge indicates that the source tool can be \emph{chained} with the destination tool: the output of tool $t_i$ can be transformed into or used as input to tool $t_j$. Since schema-level type matching alone is too coarse (e.g., two string fields may be semantically unrelated), we use an LLM to judge each candidate pair based on tool descriptions and schemas, assigning a confidence level (high, medium, or low) to each positive edge. The resulting graph contains approximately 83K directed edges (average 88 successors per tool), of which 64K are medium-or-higher confidence. The graph includes both intra-server and cross-server edges.

\subsection{Task Synthesis}
\label{sec:task_synthesis}

\subsubsection{DAG Exploration}
\label{sec:exploration}

With nearly 1,000 tools, undirected exploration would waste most calls on incoherent combinations. We use the tool graph $G$ to focus exploration on semantically valid workflows. For each exploration, we select a starting tool with at least two high-confidence successors in $G$ (enabling fan-out), then let a strong LLM interactively build a tool-call DAG against the live server. At each round, the LLM is offered a random subset of successors (from the union of all previously completed nodes in $G$, filtered to medium-or-higher confidence). The LLM may call multiple tools in parallel (fan-out), chain a single successor (sequential), or combine outputs from earlier nodes into one call (fan-in). This process repeats until a specified tool-call budget is exhausted or no successors remain.

A completed exploration yields a DAG:
\[
D = (V, E_D), \quad V = \{(t_k, a_k, r_k)\}_{k=1}^{m},
\]
where each node records tool identity $t_k$, input $a_k$, and output $r_k$, and each edge in $E_D$ indicates data flow from parent to child. Because every node results from actual execution against a live server, all states are guaranteed reachable. These DAGs provide raw material for task synthesis and are later indexed by the retrieval-augmented simulator (Sec.~\ref{sec:trajectory_generation}).

\subsubsection{Back-Chaining Task Synthesis}
\label{sec:backchain}

From each DAG $D$, the LLM selects a subset of nodes and constructs a natural-language task around them. The task embeds the information needed to populate tool-call inputs (names, parameters, constraints), while the ground-truth answer is extracted from the selected nodes' outputs. Remaining nodes serve as distractors or irrelevant exploration paths. Each generated task includes a structured \texttt{answer\_schema} (a JSON template with named placeholders) and an \texttt{answer\_template} (a natural-language sentence with the same placeholders). This structured format enables automated evaluation via field-level comparison rather than free-form text matching.

This backward construction guarantees both reachability and label correctness. To maximize diversity, we vary which subsets are selected and prompt the LLM to vary phrasing, yielding multiple tasks of different complexity from a single DAG.

\subsubsection{Task Validation}
\label{sec:validation}

An LLM judge evaluates each task--trajectory pair on five criteria:
\textbf{Verifiability} (answer is objectively checkable),
\textbf{Well-specifiedness} (exactly one correct answer),
\textbf{Interpretability} (outputs are human-readable, not opaque IDs),
\textbf{Realism} (score $\geq 5$ on a 0--10 scale), and
\textbf{Difficulty calibration} (stated difficulty matches trajectory complexity).
Tasks failing any criterion are removed. This stage filters out approximately half of candidates. Appendix~\ref{app:human_audit} reports a human audit of this filtering.

\subsection{Scalable and Verifiable Evaluation and Rollout}
\label{sec:trajectory_generation}

Real-world tools depend on external data sources that change over time (weather, stock prices, search indices), and live servers may become unavailable or rate-limited. Directly using live MCP servers for RL rollout or evaluation is therefore neither reproducible nor scalable. We construct a retrieval-augmented tool-call simulator that replaces live servers entirely, so that both RL training and evaluation run without live server access.

\subsubsection{Retrieval-Augmented Tool-Call Simulator}
\label{sec:simulator}

The simulator indexes every tool call $(t_k, a_k, r_k)$ recorded during exploration and resolves each tool call that the agent issues during RL rollout or evaluation in two stages.
In the first stage, a rule-based retriever looks up the call input in the cache.
When the input is byte-identical to a cached input, the retriever returns the cached output unchanged, and when the cache holds no similar call, the retriever rejects the call with a tool error.
Otherwise, the retriever passes the most similar cached calls to the second stage, in which an LLM resolves the call against these retrieved examples.
The LLM returns a retrieved real output unchanged when it judges the agent's call semantically identical to a cached call, for example when the two calls differ only in the plural form of a search term.
When the agent's call differs from a cached call only in a minor detail, such as requesting CSV where the cached call requested JSON, the LLM generates a new response grounded in the retrieved outputs.
When the agent's call differs semantically from every retrieved call, the LLM rejects it with a tool error.

\begin{table}[t]
\centering
\caption{Final outcomes of the retrieval-augmented simulator over 124,765 tool calls issued in 16 evaluations on the \projectname~test set, covering the base model and 15 RL checkpoints.}
\label{tab:sim_outcomes}
\small
\begin{booktabs}{
  colspec = {X c X},
  width=\linewidth,
  row{1} = {font=\bfseries},
}
\toprule
Outcome & Share of calls & Response received by the agent \\
\midrule
Retriever exact-cache return & 33.1\% & Real cached output, byte-identical \\
LLM-selected cache return & 18.4\% & Real cached output, byte-identical \\
LLM-synthesized response & 13.8\% & Generated output (median 83 characters) \\
LLM rejection & 34.7\% & Tool error \\
Retriever rejection & 0.0\% & Tool error \\
\bottomrule
\end{booktabs}
\end{table}

Table~\ref{tab:sim_outcomes} reports how often each outcome occurs during evaluation.
We verified every delivered response byte by byte against the cache of real API outputs and found that 78.9\% of the responses reaching the agent are byte-identical replays of real API outputs.
Among the synthesized responses with JSON-shaped output, 0.65\% fail to parse (34 of 5,270).
When none of the retrieved calls applies, the simulator returns an explicit error instead of inventing a response, so uncertain calls reach the agent as error signals and fabricated observations stay out of the rollout.

During retrieval, the cached calls from the current task's ground-truth trajectory are ranked ahead of other cached calls.
The prioritization keeps the simulator consistent with the world state from which each label was extracted, which matters since many tools are non-deterministic.
Weather and stock tools, for example, return different outputs for the same input over time, and some tools, such as fortune-telling tools, return a random output on every call.
Each ground-truth answer was extracted from one specific observed response, so without the prioritization such tasks would no longer be verifiable against their own labels.
To check that the prioritization does not leak answers, we used Claude Sonnet 4.5 to classify every evaluation tool call by its deviation from the cached calls.
In successful trajectories, 75.8\% of calls exactly match the ground-truth calls, whereas in failed trajectories, 66.8\% of calls deviate majorly.
If the prioritization leaked answers, calls with major deviations would succeed spuriously, and the observed pattern shows the opposite.

The simulator design guarantees two properties. \emph{Reproducibility}: tool responses are either replayed from the cache or grounded in it, so they do not depend on the time of the call or on the state of live servers.
\emph{Verifiability}: ground-truth labels were extracted from exploration outputs that are indexed in the simulator, so the responses needed to derive each label are always available via exact match.

\subsubsection{RL Rollout and Evaluation}

During RL training, the agent interacts with the simulator rather than live servers. Given a task description and tool schemas, the agent generates tool calls that are intercepted and resolved by the simulator. The binary reward (correct/incorrect) is computed by first attempting exact field-level match against the structured \texttt{answer\_schema}; if that fails, an LLM judge evaluates semantic equivalence. The same simulator serves as the evaluation environment, ensuring that training and evaluation use identical tool behavior. The simulator therefore enables unlimited rollouts at scale without API costs or availability concerns.
Since the policy collects fresh rollouts from the simulator at every training step, the learning setting is online RL, and only the environment is cached. We therefore refer to the training as simulator-based online RL.

\section{Dataset Analysis}
\label{sec:dataset}

We apply the full pipeline to construct \projectname, a large-scale dataset of verifiable tool-calling tasks grounded in real MCP servers. Starting from approximately 3,000 servers discovered on the Smithery registry, the filtering stage retains 240 servers exposing 993 tools. Tool-chain exploration produces 9,749 DAG trajectories, from which back-chaining synthesis generates 9,275 candidate tasks. After LLM-based validation, 5,144 tasks pass all quality criteria. We hold out 200 tasks as a fixed test set and use the remaining 4,944 for training. Each task is accompanied by its ground-truth trajectory, answer schema, and the full set of tool-call responses needed for offline replay.

\paragraph{Example task.}
As a concrete illustration, consider the following task from \projectname:

\begin{quote}
\textit{``Which domain was registered first, amazon.com or netflix.com, and how many years apart were they registered?''}
\end{quote}

\noindent The ground-truth trajectory makes two calls to a \texttt{whois\_lookup} tool (from the \texttt{networkcalc-mcp} server): one for \texttt{amazon.com} and one for \texttt{netflix.com}. The expected answer is:

\begin{verbatim}
{"first_registered_domain": "amazon.com",
 "amazon_registration_year": "1994",
 "netflix_registration_year": "1997",
 "years_apart": "3"}
\end{verbatim}

\noindent This task was synthesized by back-chaining from the two outputs observed during exploration, and is verifiable because the simulator can replay the exact same responses at evaluation time.

\paragraph{Task characteristics.}

Tasks span three difficulty levels: easy (19.5\%), medium (77.0\%), and hard (3.5\%). Ground-truth trajectories require on average 3.0 tool calls (range 1--10), with answer schemas containing 4.6 fields on average, reflecting multi-faceted information needs rather than single-value lookups. The test set alone covers 98 distinct servers and 221 distinct tools. Approximately 38.5\% of tasks require tools from multiple servers, exercising cross-server composition, a capability absent from most existing benchmarks that confine tasks to a single API provider. The average realism score assigned by the LLM judge is 7.8/10, indicating that the majority of tasks reflect plausible real-world use cases rather than contrived synthetic scenarios.

The retrieval-augmented simulator indexes all tool calls from the exploration DAGs. During the full training run, the retriever returns 42.2\% of agent tool calls from the exact cache and passes the remaining 57.8\% to the LLM stage, and it rejects no call for lack of similar cached calls. The LLM stage can still reject a call, and Table~\ref{tab:sim_outcomes} reports the final outcomes during evaluation, including the 34.7\% of calls that the LLM rejects. The absence of retriever rejections shows that the exploration phase covers every tool that the agent calls during training, so trajectories can be collected without live server access.

The entire dataset is generated using Claude Sonnet 4.5 via AWS Bedrock batch inference. The full pipeline consumes approximately 23.5 billion tokens (21.4B input, 2.0B output) at a total cost of approximately \$47K. This one-time investment produces a reusable dataset and simulator that enable unlimited simulator-based RL training without further API costs.

\subsection{Comparison with Existing Datasets}

\begin{table}[t]
\centering
\caption{Comparison of \projectname~with existing tool-calling datasets.}
\label{tab:comparison}
\small
\begin{booktabs}{
  colspec = {X c c c c c},
  width=\linewidth,
  row{1} = {font=\bfseries},
}
\toprule
Dataset & Tasks & Servers/APIs & Multi-step & Verifiable & Real APIs \\
\midrule
API-Bank & 1,888 & 2,138 & \checkmark & \checkmark & $\times$ \\
ToolAlpaca & 3,938 & 400+ & $\times$ & $\times$ & \checkmark \\
APIGen-MT & 5,000 & 28 & \checkmark & \checkmark & $\times$ \\
Tau2-Bench & 279 & 68 & \checkmark & \checkmark & $\times$ \\
Toucan & 1.5M & 2000+ & \checkmark & $\times$ & \checkmark \\
\midrule
\projectname & 5,144 & 993 & \checkmark & \checkmark & \checkmark \\
\bottomrule
\end{booktabs}
\end{table}

Table~\ref{tab:comparison} highlights the key differentiators of \projectname: it is the only dataset that simultaneously provides multi-step trajectories, verifiable ground-truth labels, and grounding in real (not simulated) API servers. Existing datasets either rely on simulated environments (APIGen-MT, Tau2-Bench), lack verifiable supervision (Toucan), or are limited to single-turn interactions (ToolAlpaca).

\section{Experiments}
\label{sec:experiments}

We evaluate whether training on \projectname~improves tool-calling ability. We fine-tune a model using reinforcement learning and evaluate on both our held-out test set and public benchmarks.

\subsection{Setup}

\paragraph{Model.}
We train \textbf{Qwen3-4B} \cite{yangQwen3TechnicalReport2025}, starting from Qwen3-4B-Thinking-2507.

\paragraph{Training.}
We use GRPO~\citep{shaoDeepSeekMathPushingLimits2024} with dynamic filtering~\citep{yuDAPOOpenSourceLLM2025} on the 4,944 training tasks from \projectname. During RL, the agent interacts with the retrieval-augmented simulator: at each step it selects and calls tools, receives simulated responses, and produces a final answer.
We use a binary reward: we first attempt structured field-level exact matching against the \texttt{answer\_schema}; if that fails, an LLM judge (Qwen3-30B-A3B-Thinking-2507) evaluates semantic equivalence.
The same model also serves as the LLM backend for the tool-call simulator.
Appendix~\ref{app:swap} shows that the training curve and the RL gain are preserved when both roles are taken over by Qwen3-235B-A22B or by Claude Sonnet 4.5.

We sample 8 rollouts per prompt with temperature 1.0 and apply dynamic filtering~\citep{yuDAPOOpenSourceLLM2025}. Prompts where the model achieves reward 1.0 on all 8 rollouts are \emph{permanently removed} from the training set, as the model has already mastered them. To avoid premature exclusion early in training when rewards may be noisy, we only begin removing mastered prompts once more than 10 accumulate within a single rollout batch. Prompts where all rollouts fail are \emph{not} excluded, as the model may solve them after further training. This curriculum-like mechanism progressively focuses training on the shrinking set of prompts the model finds challenging.

We train for 300 steps with a global batch size of 16 ($\approx$5 epochs with filtering). Key hyperparameters are provided in Appendix~\ref{app:hyperparams}. Training uses the Slime framework with 16 GPUs for the actor model and 16 GPUs for the asynchronous rollout engine.

\paragraph{Evaluation.}
We evaluate on four benchmark suites: \textbf{\projectname~test set} (200 held-out tasks; pass@$k$ for $k \in \{1, 4, 8, 16\}$ over 16 independent rollouts per task), \textbf{Tau2-Bench}~\citep{yaotaubenchBenchmarkToolAgentUser2024,barrestau2BenchEvaluatingConversational2025} (Retail, Airline, and Telecom), \textbf{MCP-Atlas}, and \textbf{MCPMark} (File System, Postgres Easy, and Postgres Std).
For the \projectname~test set, we use the offline simulator with all DAG tools available and an LLM judge for answer comparison; we also evaluate Claude Haiku 4.5, Sonnet 4.6, and Opus 4.7 under the same protocol as proprietary baselines.

\subsection{Results}

\subsubsection{\projectname~Test Set}

\begin{table}[t]
\centering
\caption{Performance on the \projectname~test set. We compare our fine-tuned Qwen3-4B against its base checkpoint and proprietary Claude models evaluated under identical conditions.}
\label{tab:firefly_results}
\begin{booktabs}{
  colspec = {X c c c c},
  width=\linewidth,
  row{1} = {font=\bfseries},
}
\toprule
Model & Pass@1 & Pass@4 & Pass@8 & Pass@16 \\
\midrule
Qwen3-4B (base) & 28.1\% & 34.9\% & 38.4\% & 43.0\% \\
Claude Haiku 4.5 & 30.0\% & 43.2\% & 48.7\% & 54.0\% \\
Claude Sonnet 4.6 & 42.2\% & 48.3\% & 50.3\% & 53.0\% \\
Claude Opus 4.7 & 44.5\% & 51.7\% & 54.4\% & 57.0\% \\
\midrule
Qwen3-4B + \projectname~RL & {41.5\%} & {49.3\%} & {52.8\%} & {57.0\%} \\
\bottomrule
\end{booktabs}
\end{table}

Table~\ref{tab:firefly_results} presents results on the \projectname~test set. RL training on \projectname~improves Qwen3-4B from 28.1\% to 41.5\% pass@1 (+13.4 points), matching Claude Sonnet 4.6 (42.2\%) despite being orders of magnitude smaller. At pass@8, the fine-tuned 4B model reaches 52.8\%, surpassing Claude Sonnet (50.3\%) and approaching Claude Opus (54.4\%).

As shown in Figure~\ref{fig:training_curve} in Appendix~\ref{app:more_results}, the model improves steadily over training. The pass@16 reaches 57.0\%, a modest gap over pass@1 (41.5\%), indicating that some harder tasks remain beyond the 4B model's capacity even with multiple attempts. We hypothesize that scaling to larger models would substantially increase this ceiling (see Sec.~\ref{sec:limitations}).

\subsubsection{Public Benchmarks}

\begin{table}[t]
\centering
\caption{Results on public benchmarks. \projectname~RL improves Qwen3-4B-Thinking-2507 across all evaluated benchmarks.}
\label{tab:public_benchmarks}
\begin{booktabs}{
  colspec = {Q[l,wd=.4\linewidth] X[c] X[c]X[c]X[c]},
  width={\linewidth},
  row{1} = {font=\bfseries},
  cells={m},
}
\toprule
Model & {Tau2-Bench\\Retail} & {Tau2-Bench\\Airline} & {Tau2-Bench\\Telecom} & MCP-Atlas \\
\midrule
Qwen3-4B & 0.491 & 0.365 & 0.189 & 19.4\% \\
Qwen3-4B + \projectname~RL & \textbf{0.627} & \textbf{0.525} & \textbf{0.204} & \textbf{26.0\%} \\
\bottomrule
\end{booktabs}

\vspace{1em}

\begin{booktabs}{
  colspec = {Q[l,wd=.4\linewidth] X[c]X[c]X[c]},
  width={\linewidth},
  row{1} = {font=\bfseries},
  cells={m},
}
\toprule
Model & {MCPMark\\File System} & {MCPMark\\Postgres Easy} & {MCPMark\\Postgres Std} \\
\midrule
Qwen3-4B & 40.0\% & 70.0\% & 9.5\% \\
Qwen3-4B + \projectname~RL & \textbf{60.0\%} & \textbf{80.0\%} & \textbf{13.3\%} \\
\bottomrule
\end{booktabs}
\end{table}

\projectname is constructed as a single-turn, multi-step tool-calling dataset: each task contains one user message, but solving it requires the agent to plan and execute multiple tool calls before producing the final answer.
We evaluate whether supervision from this format transfers to public benchmarks that differ substantially from \projectname in tool source and environment design.
Specifically, MCPMark and MCP-Atlas evaluate tool use on MCP servers outside \projectname, whereas Tau2-Bench evaluates multi-turn user-agent interactions.

As shown in Table~\ref{tab:public_benchmarks}, \projectname~RL improves Qwen3-4B across all evaluated public benchmarks.
On Tau2-Bench, the model improves from 0.491 to 0.627 on Retail, from 0.365 to 0.525 on Airline, and from 0.189 to 0.204 on Telecom.
On MCP-Atlas, accuracy improves from 19.4\% to 26.0\%.
On MCPMark, the model also improves across all evaluated subsets, from 40.0\% to 60.0\% on File System, from 70.0\% to 80.0\% on Postgres Easy, and from 9.5\% to 13.3\% on Postgres Standard.

These consistent gains suggest that the model does not merely memorize cached trajectories from \projectname.
Instead, training on verified single-turn multi-step tasks appears to improve transferable tool-use behaviors, including selecting relevant tools, grounding arguments in observations, chaining intermediate outputs, and deriving final answers from structured tool responses.
The improvements on MCPMark show transfer to real MCP execution environments, while the gains on MCP-Atlas show transfer to manually designed MCP-style servers.
Most notably, the improvements on Tau2-Bench indicate that single-turn multi-step supervision can also benefit multi-turn tool use: although Tau2-Bench requires interaction across multiple user turns, many of its core difficulties still involve robust tool selection, state tracking, API grounding, and multi-step reasoning.
Overall, these results support the claim that \projectname provides broadly useful supervision for tool-calling agents beyond its own data distribution.

To compare RL with supervised learning on the same tasks, we also fine-tune the same Qwen3-4B base model on trajectories that Claude Sonnet 4.5 generates for the training tasks and that reach correct answers.
Supervised fine-tuning on these Claude-generated samples improves pass@1 on the \projectname~test set from 28.1\% to 38.5\% after three epochs, and RL on \projectname~reaches 41.5\%, outperforming the best supervised configuration by 3.0 points (Table~\ref{tab:sft_baseline} in Appendix~\ref{app:more_results}).
Appendix~\ref{app:more_results} also compares our model with models trained on the Toucan dataset~\citep{xuTOUCANSynthesizing15M2025}, and our model achieves higher scores on all three Tau2-Bench domains.

\section{Conclusion}

We presented \projectname, a pipeline for generating large-scale, verified tool-call training data from real-world MCP servers.
By inverting the standard synthesis pipeline (explore first, then back-chain tasks from observed outcomes), \projectname makes label correctness a property of data generation rather than post-hoc filtering.
A pairwise tool graph scales exploration to nearly 1,000 tools, while a retrieval-augmented simulator enables reproducible simulator-based online RL training and evaluation despite environment drift in live APIs.
The resulting dataset of 5,144 verified tasks across 240 servers enables a 4B model to match Claude Sonnet 4.6 on our held-out test set, and the same model improves on multiple out-of-distribution tool-calling benchmarks, although our comparison with Claude Sonnet 4.6 is limited to the held-out test set.
We release the code at \url{https://github.com/leoleoasd/firefly-neurips-2026/} and the data, simulator, and model at \url{http://firefly.hailab.io/} to support scalable, verifiable agent learning.

\section{Limitations}
\label{sec:limitations}

\paragraph{Model scale.}
Due to computational constraints, we only train and evaluate a 4B-parameter model. While the results demonstrate clear gains from \projectname's verifiable supervision, larger models are likely to benefit even more: the modest gap between pass@1 and pass@16 (41.5\% vs.\ 57.0\%) suggests that the 4B model lacks the capacity to solve many harder multi-step tasks regardless of sampling budget. Scaling to larger models (e.g., 30B+ parameters) would likely raise both the pass@1 ceiling and the pass@16 coverage, as larger models can internalize more diverse solution strategies from the training trajectories.

\paragraph{Single-turn interaction.}
All tasks in \projectname~follow a single-turn format: the user issues one request and the agent completes it through multi-step tool calls. Real-world tool-use scenarios often involve multi-turn conversations where the user provides clarifications, changes requirements, or builds on previous results. Extending the pipeline to generate and validate multi-turn dialogues with interleaved user messages is an important direction for future work.

\paragraph{Conservative simulator errors.}
The simulator returns a tool error for calls that differ semantically from every cached call, so an agent that finds a valid path outside the explored trajectories during RL receives an error for that path, which may narrow the policy toward explored paths.
To bound this truncation risk, we classified every evaluation tool call as an exact, minor, or major deviation from the nearest cached call and related the classes to task success.
In successful trajectories, about 24\% of calls deviate from the cached ground-truth path (13.8\% with major and 10.4\% with minor deviations) and still lead to full reward, so alternative paths can pass through the simulator.
In failed trajectories, 66.8\% of calls deviate majorly, so the errors fall mostly on semantically wrong calls, such as calls with wrong entities or query scope, and seldom on valid reformulations.
After an error, the agent can retry with corrected arguments within its round budget, as it would after a real API returns a client error.
Moreover, the untrained base model shows the same relation between deviation and failure, so the errors do not selectively shape the trained policy.

\paragraph{Generation cost.}
Generating \projectname~costs approximately \$47K in API usage (Sec.~\ref{sec:dataset}), which may limit how easily smaller teams can regenerate the dataset around their own tools.
The total amounts to about \$9 per verified task, and each task comes with a ground-truth trajectory, an answer schema, and a replayable environment.
Since we release the dataset, the simulator, and the trained model, downstream users can train and evaluate without API costs, and teams with proprietary tools can start from the released model instead of regenerating a dataset.

\paragraph{Number of servers.}
\projectname~covers 240 servers because the four filtering criteria in Sec.~\ref{sec:filtering} retain only 240 of roughly 3,000 scraped servers, so the supply of qualifying public servers limits the dataset's scale more than the pipeline does.
The pipeline itself is registry-agnostic, and new servers can be added incrementally.
New tools join the graph through pairwise judgments against existing nodes, and exploration extends to their neighborhoods without re-processing existing servers.

\clearpage
\begin{ack}
This work received no third-party funding. The authors declare no competing interests.
\end{ack}

\bibliography{custom}
\bibliographystyle{plainnat}

\appendix
\section{Training Hyperparameters}
\label{app:hyperparams}

\begin{table}[h]
\centering
\caption{Training hyperparameters}
\label{tab:hyperparams}
\small
\begin{booktabs}{
  colspec = {X l},
  width=0.6\linewidth,
  row{1} = {font=\bfseries},
}
\toprule
Parameter & Value \\
\midrule
Learning rate & $10^{-6}$ (constant) \\
PPO clip $\epsilon$ & 0.2 \\
KL loss coefficient & 0.001 \\
Weight decay & 0.01 \\
Adam $\beta_1, \beta_2$ & 0.9, 0.98 \\
Max response length & 4096 tokens \\
Rollouts per prompt & 8 \\
Temperature & 1.0 \\
Global batch size & 16 \\
Training steps & 300 \\
\bottomrule
\end{booktabs}
\end{table}

\section{Additional Results}
\label{app:more_results}

\begin{figure}[h]
    \centering
    \includegraphics[width=\textwidth]{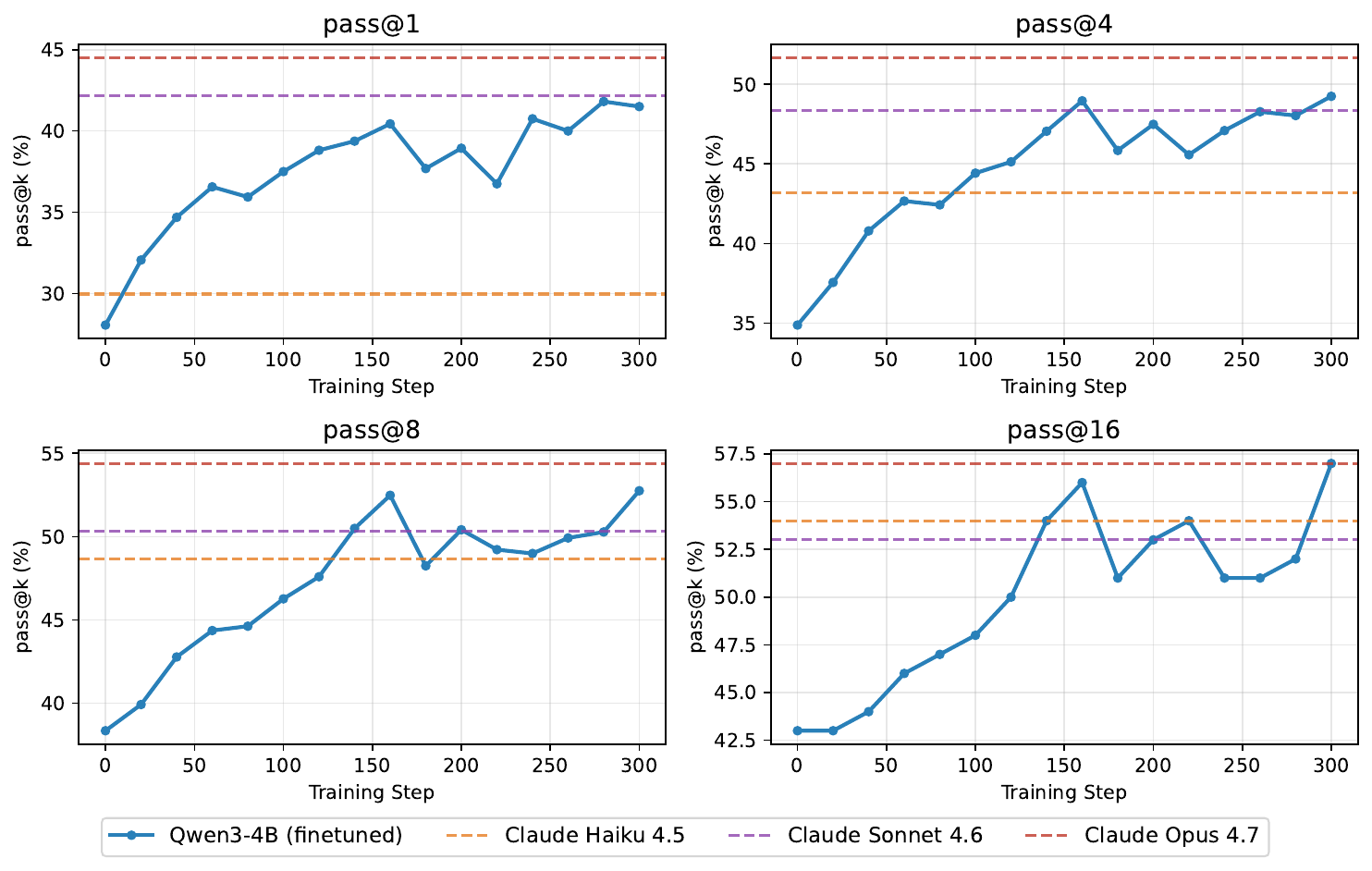}
    \caption{Pass@$k$ on the \projectname~test set over training. The model improves steadily across all $k$ values throughout RL training.}
    \label{fig:training_curve}
\end{figure}

Figure~\ref{fig:training_curve} shows pass@$k$ on the \projectname~test set over RL training.

\paragraph{Comparison with supervised fine-tuning.}
To compare RL with supervised learning on the same tasks, we fine-tune the same Qwen3-4B base model on samples generated by Claude Sonnet 4.5.
We sample four trajectories from Claude Sonnet 4.5 on each training task, keep the trajectories whose final answers are correct, and fine-tune the base model on them.
As shown in Table~\ref{tab:sft_baseline}, supervised fine-tuning on these Claude-generated samples already improves pass@1 from 28.1\% to 38.5\% after three epochs, which indicates that the \projectname~tasks provide useful supervision under both training paradigms.
RL on \projectname~still outperforms the best supervised configuration by 3.0 points of pass@1.

\paragraph{Comparison with prior synthetic data.}
To compare with prior synthetic data built on real MCP servers, we take the Tau2-Bench results reported for models trained on Toucan~\citep{xuTOUCANSynthesizing15M2025}, which generates tool-use queries first and then has teacher models produce trajectories over real MCP servers.
Toucan trains Qwen2.5-Instruct models, whereas we train Qwen3-4B-Thinking-2507, so Table~\ref{tab:toucan} also reports each method's improvement over its own base model.
Our 4B model achieves higher scores than Toucan-trained models of up to 32B parameters on all three Tau2-Bench domains, and it also obtains the largest improvement over its own base model.

\begin{table}[h]
\centering
\caption{Supervised fine-tuning on Claude-generated samples versus RL on the \projectname~test set. All rows use the same base model and evaluation protocol.}
\label{tab:sft_baseline}
\small
\begin{booktabs}{
  colspec = {X c c},
  width=\linewidth,
  row{1} = {font=\bfseries},
}
\toprule
Model & Pass@1 & Pass@16 \\
\midrule
Qwen3-4B (base) & 28.1\% & 43.0\% \\
+ SFT on Claude-generated samples, 1 epoch & 35.9\% & 53.0\% \\
+ SFT on Claude-generated samples, 3 epochs & 38.5\% & 56.0\% \\
+ \projectname~RL & 41.5\% & 57.0\% \\
\bottomrule
\end{booktabs}
\end{table}

\begin{table}[h]
\centering
\caption{Tau2-Bench scores of models trained on Toucan, as reported by \citet{xuTOUCANSynthesizing15M2025}, and of our model trained on \projectname. Scores are multiplied by 100. The base models differ, so the last column reports each method's improvement over its own base model.}
\label{tab:toucan}
\small
\begin{booktabs}{
  colspec = {X c c c c c c},
  width=\linewidth,
  row{1} = {font=\bfseries},
}
\toprule
Model & Size & Retail & Airline & Telecom & Avg. & {$\Delta$ over\\base} \\
\midrule
Toucan-Qwen2.5-7B & 7B & 22.8 & 20.0 & 10.5 & 17.8 & +1.7 \\
Toucan-Qwen2.5-14B & 14B & 49.1 & 22.0 & 20.2 & 30.4 & +6.0 \\
Toucan-Qwen2.5-32B & 32B & 52.6 & 22.0 & 20.2 & 31.6 & +2.2 \\
\midrule
Qwen3-4B + \projectname~RL & 4B & 62.7 & 52.5 & 20.4 & 45.2 & +10.4 \\
\bottomrule
\end{booktabs}
\end{table}

\section{Robustness to the Judge and Simulator Model}
\label{app:swap}

During training and in the main evaluation, Qwen3-30B-A3B-Thinking-2507 serves as both the reward judge and the LLM backend of the simulator, so the trained policy could in principle adapt to this specific model.
To test this possibility, we re-evaluate all 15 RL checkpoints twice, each time replacing both the judge and the simulator backend, once with Qwen3-235B-A22B from the same model family and once with Claude Sonnet 4.5 from a different model family.
Table~\ref{tab:swap} reports pass@1 on the \projectname~test set under the three settings.
Both replacements preserve the shape of the training curve and the end-to-end RL gain, and all three settings place the pass@1 peak at the same checkpoint.
The Qwen3-235B-A22B setting shifts pass@1 by $-0.96$ points on average with no systematic difference at pass@16, whereas the Claude Sonnet 4.5 setting shifts pass@1 upward by 4.2 points on average, with a Pearson correlation of 0.961 against the original curve.
If the original judge or simulator backend distorted the outcomes, replacing it with a much stronger model would change the curve beyond such a near-uniform shift.

\begin{table}[h]
\centering
\caption{Pass@1 (\%) of each RL checkpoint on the \projectname~test set when the reward judge and the simulator backend are both Qwen3-30B-A3B-Thinking-2507 (the setting used in the main text), both Qwen3-235B-A22B, or both Claude Sonnet 4.5.}
\label{tab:swap}
\small
\begin{booktabs}{
  colspec = {X c c c},
  width=\linewidth,
  row{1} = {font=\bfseries},
}
\toprule
RL step & Qwen3-30B (main) & Qwen3-235B & Claude Sonnet 4.5 \\
\midrule
19 & 32.1 & 31.1 & 36.1 \\
39 & 34.7 & 33.6 & 39.3 \\
59 & 36.6 & 34.4 & 40.8 \\
79 & 35.9 & 35.6 & 39.9 \\
99 & 37.5 & 36.4 & 41.5 \\
119 & 38.8 & 38.5 & 44.2 \\
139 & 39.4 & 37.7 & 44.0 \\
159 & 40.4 & 38.2 & 43.9 \\
179 & 37.7 & 37.8 & 43.5 \\
199 & 38.9 & 38.9 & 43.8 \\
219 & 36.8 & 36.1 & 40.9 \\
239 & 40.8 & 39.8 & 44.4 \\
259 & 40.0 & 40.0 & 43.4 \\
279 & 41.8 & 40.7 & 45.4 \\
299 & 41.5 & 39.8 & 44.8 \\
\bottomrule
\end{booktabs}
\end{table}

\section{Human Audit of Task Validation}
\label{app:human_audit}

To measure how well the LLM judge in Sec.~\ref{sec:validation} filters tasks, we randomly sampled 50 tasks that passed the judge and 50 candidates that it rejected, and human annotators judged these 100 tasks.
The two groups were interleaved in random order, and the annotators did not know which group each task came from.
Each task was judged on three binary criteria and one rating.
Instruction clarity asks whether the task text together with the tool input schema suffices to construct the tool call.
Answer uniqueness asks whether exactly one answer is defensible.
Label correctness asks whether every field of the recorded answer is supported by a tool output and not only by a tool input.
Realism is rated on a scale from 1 to 5.

\begin{table}[h]
\centering
\caption{Human audit of 50 tasks that passed and 50 candidates that were rejected by the LLM judge. The fifth column gives the share of tasks that satisfy all three binary criteria, and realism is the mean rating on a 1--5 scale.}
\label{tab:human_audit}
\small
\begin{booktabs}{
  colspec = {X c c c c c},
  width=\linewidth,
  row{1} = {font=\bfseries},
}
\toprule
Judge decision & Clarity & Uniqueness & Label correctness & All three & Realism \\
\midrule
Passed ($n=50$) & 84.0\% & 82.0\% & 80.0\% & 78.0\% & 4.50 \\
Rejected ($n=50$) & 64.0\% & 66.0\% & 72.0\% & 38.0\% & 4.28 \\
\bottomrule
\end{booktabs}
\end{table}

As shown in Table~\ref{tab:human_audit}, tasks that passed the judge satisfy all three binary criteria 78.0\% of the time, compared with 38.0\% for rejected candidates, and the difference is statistically significant ($\chi^2 = 16.4$, $\mathrm{df} = 1$, $p < 0.001$).

\section*{NeurIPS Paper Checklist}

\begin{enumerate}

\item {\bf Claims}
    \item[] Question: Do the main claims made in the abstract and introduction accurately reflect the paper's contributions and scope?
    \item[] Answer: \answerYes{}
    \item[] Justification: The abstract and introduction clearly state three contributions (dataset, framework, experimental results) that are supported by the method description and experiments.
    \item[] Guidelines:
    \begin{itemize}
        \item The answer \answerNA{} means that the abstract and introduction do not include the claims made in the paper.
        \item The abstract and/or introduction should clearly state the claims made, including the contributions made in the paper and important assumptions and limitations. A \answerNo{} or \answerNA{} answer to this question will not be perceived well by the reviewers. 
        \item The claims made should match theoretical and experimental results, and reflect how much the results can be expected to generalize to other settings. 
        \item It is fine to include aspirational goals as motivation as long as it is clear that these goals are not attained by the paper. 
    \end{itemize}

\item {\bf Limitations}
    \item[] Question: Does the paper discuss the limitations of the work performed by the authors?
    \item[] Answer: \answerYes{}
    \item[] Justification: Sec.~\ref{sec:limitations} discusses model scale (only 4B trained due to compute), the single-turn interaction format, conservative simulator errors, generation cost, and the number of servers.
    \item[] Guidelines:
    \begin{itemize}
        \item The answer \answerNA{} means that the paper has no limitation while the answer \answerNo{} means that the paper has limitations, but those are not discussed in the paper. 
        \item The authors are encouraged to create a separate ``Limitations'' section in their paper.
        \item The paper should point out any strong assumptions and how robust the results are to violations of these assumptions (e.g., independence assumptions, noiseless settings, model well-specification, asymptotic approximations only holding locally). The authors should reflect on how these assumptions might be violated in practice and what the implications would be.
        \item The authors should reflect on the scope of the claims made, e.g., if the approach was only tested on a few datasets or with a few runs. In general, empirical results often depend on implicit assumptions, which should be articulated.
        \item The authors should reflect on the factors that influence the performance of the approach. For example, a facial recognition algorithm may perform poorly when image resolution is low or images are taken in low lighting. Or a speech-to-text system might not be used reliably to provide closed captions for online lectures because it fails to handle technical jargon.
        \item The authors should discuss the computational efficiency of the proposed algorithms and how they scale with dataset size.
        \item If applicable, the authors should discuss possible limitations of their approach to address problems of privacy and fairness.
        \item While the authors might fear that complete honesty about limitations might be used by reviewers as grounds for rejection, a worse outcome might be that reviewers discover limitations that aren't acknowledged in the paper. The authors should use their best judgment and recognize that individual actions in favor of transparency play an important role in developing norms that preserve the integrity of the community. Reviewers will be specifically instructed to not penalize honesty concerning limitations.
    \end{itemize}

\item {\bf Theory assumptions and proofs}
    \item[] Question: For each theoretical result, does the paper provide the full set of assumptions and a complete (and correct) proof?
    \item[] Answer: \answerNA{}
    \item[] Justification: The paper does not include theoretical results or proofs.
    \item[] Guidelines:
    \begin{itemize}
        \item The answer \answerNA{} means that the paper does not include theoretical results. 
        \item All the theorems, formulas, and proofs in the paper should be numbered and cross-referenced.
        \item All assumptions should be clearly stated or referenced in the statement of any theorems.
        \item The proofs can either appear in the main paper or the supplemental material, but if they appear in the supplemental material, the authors are encouraged to provide a short proof sketch to provide intuition. 
        \item Inversely, any informal proof provided in the core of the paper should be complemented by formal proofs provided in appendix or supplemental material.
        \item Theorems and Lemmas that the proof relies upon should be properly referenced. 
    \end{itemize}

    \item {\bf Experimental result reproducibility}
    \item[] Question: Does the paper fully disclose all the information needed to reproduce the main experimental results of the paper to the extent that it affects the main claims and/or conclusions of the paper (regardless of whether the code and data are provided or not)?
    \item[] Answer: \answerYes{}
    \item[] Justification: We provide the full pipeline description, training hyperparameters (Appendix~\ref{app:hyperparams}), dataset, and evaluation setup. Code is available at \url{https://github.com/leoleoasd/firefly-neurips-2026/}, and data and model are available at \url{http://firefly.hailab.io/}.
    \item[] Guidelines:
    \begin{itemize}
        \item The answer \answerNA{} means that the paper does not include experiments.
        \item If the paper includes experiments, a \answerNo{} answer to this question will not be perceived well by the reviewers: Making the paper reproducible is important, regardless of whether the code and data are provided or not.
        \item If the contribution is a dataset and\slash or model, the authors should describe the steps taken to make their results reproducible or verifiable. 
        \item Depending on the contribution, reproducibility can be accomplished in various ways. For example, if the contribution is a novel architecture, describing the architecture fully might suffice, or if the contribution is a specific model and empirical evaluation, it may be necessary to either make it possible for others to replicate the model with the same dataset, or provide access to the model. In general. releasing code and data is often one good way to accomplish this, but reproducibility can also be provided via detailed instructions for how to replicate the results, access to a hosted model (e.g., in the case of a large language model), releasing of a model checkpoint, or other means that are appropriate to the research performed.
        \item While NeurIPS does not require releasing code, the conference does require all submissions to provide some reasonable avenue for reproducibility, which may depend on the nature of the contribution. For example
        \begin{enumerate}
            \item If the contribution is primarily a new algorithm, the paper should make it clear how to reproduce that algorithm.
            \item If the contribution is primarily a new model architecture, the paper should describe the architecture clearly and fully.
            \item If the contribution is a new model (e.g., a large language model), then there should either be a way to access this model for reproducing the results or a way to reproduce the model (e.g., with an open-source dataset or instructions for how to construct the dataset).
            \item We recognize that reproducibility may be tricky in some cases, in which case authors are welcome to describe the particular way they provide for reproducibility. In the case of closed-source models, it may be that access to the model is limited in some way (e.g., to registered users), but it should be possible for other researchers to have some path to reproducing or verifying the results.
        \end{enumerate}
    \end{itemize}

\item {\bf Open access to data and code}
    \item[] Question: Does the paper provide open access to the data and code, with sufficient instructions to faithfully reproduce the main experimental results, as described in supplemental material?
    \item[] Answer: \answerYes{}
    \item[] Justification: Code is available at \url{https://github.com/leoleoasd/firefly-neurips-2026/}, and data and model are available at \url{http://firefly.hailab.io/}.
    \item[] Guidelines:
    \begin{itemize}
        \item The answer \answerNA{} means that paper does not include experiments requiring code.
        \item Please see the NeurIPS code and data submission guidelines (\url{https://neurips.cc/public/guides/CodeSubmissionPolicy}) for more details.
        \item While we encourage the release of code and data, we understand that this might not be possible, so \answerNo{} is an acceptable answer. Papers cannot be rejected simply for not including code, unless this is central to the contribution (e.g., for a new open-source benchmark).
        \item The instructions should contain the exact command and environment needed to run to reproduce the results. See the NeurIPS code and data submission guidelines (\url{https://neurips.cc/public/guides/CodeSubmissionPolicy}) for more details.
        \item The authors should provide instructions on data access and preparation, including how to access the raw data, preprocessed data, intermediate data, and generated data, etc.
        \item The authors should provide scripts to reproduce all experimental results for the new proposed method and baselines. If only a subset of experiments are reproducible, they should state which ones are omitted from the script and why.
        \item At submission time, to preserve anonymity, the authors should release anonymized versions (if applicable).
        \item Providing as much information as possible in supplemental material (appended to the paper) is recommended, but including URLs to data and code is permitted.
    \end{itemize}

\item {\bf Experimental setting/details}
    \item[] Question: Does the paper specify all the training and test details (e.g., data splits, hyperparameters, how they were chosen, type of optimizer) necessary to understand the results?
    \item[] Answer: \answerYes{}
    \item[] Justification: Training details including optimizer, learning rate, batch size, and all hyperparameters are provided in Sec.~\ref{sec:experiments} and Appendix~\ref{app:hyperparams}. Data splits (4,944 train / 200 test) are specified in Sec.~\ref{sec:dataset}.
    \item[] Guidelines:
    \begin{itemize}
        \item The answer \answerNA{} means that the paper does not include experiments.
        \item The experimental setting should be presented in the core of the paper to a level of detail that is necessary to appreciate the results and make sense of them.
        \item The full details can be provided either with the code, in appendix, or as supplemental material.
    \end{itemize}

\item {\bf Experiment statistical significance}
    \item[] Question: Does the paper report error bars suitably and correctly defined or other appropriate information about the statistical significance of the experiments?
    \item[] Answer: \answerNo{}
    \item[] Justification: We report pass@$k$ for $k \in \{1, 4, 8, 16\}$ computed over 16 independent rollouts per task, which captures sampling variance. The training curve (Figure~\ref{fig:training_curve}) shows stability across passes.
    \item[] Guidelines:
    \begin{itemize}
        \item The answer \answerNA{} means that the paper does not include experiments.
        \item The authors should answer \answerYes{} if the results are accompanied by error bars, confidence intervals, or statistical significance tests, at least for the experiments that support the main claims of the paper.
        \item The factors of variability that the error bars are capturing should be clearly stated (for example, train/test split, initialization, random drawing of some parameter, or overall run with given experimental conditions).
        \item The method for calculating the error bars should be explained (closed form formula, call to a library function, bootstrap, etc.)
        \item The assumptions made should be given (e.g., Normally distributed errors).
        \item It should be clear whether the error bar is the standard deviation or the standard error of the mean.
        \item It is OK to report 1-sigma error bars, but one should state it. The authors should preferably report a 2-sigma error bar than state that they have a 96\% CI, if the hypothesis of Normality of errors is not verified.
        \item For asymmetric distributions, the authors should be careful not to show in tables or figures symmetric error bars that would yield results that are out of range (e.g., negative error rates).
        \item If error bars are reported in tables or plots, the authors should explain in the text how they were calculated and reference the corresponding figures or tables in the text.
    \end{itemize}

\item {\bf Experiments compute resources}
    \item[] Question: For each experiment, does the paper provide sufficient information on the computer resources (type of compute workers, memory, time of execution) needed to reproduce the experiments?
    \item[] Answer: \answerYes{}
    \item[] Justification: We report GPU counts (16 actor + 16 rollout) and dataset generation cost (\$47K, 23.5B tokens) in Secs.~\ref{sec:dataset} and~\ref{sec:experiments}.
    \item[] Guidelines:
    \begin{itemize}
        \item The answer \answerNA{} means that the paper does not include experiments.
        \item The paper should indicate the type of compute workers CPU or GPU, internal cluster, or cloud provider, including relevant memory and storage.
        \item The paper should provide the amount of compute required for each of the individual experimental runs as well as estimate the total compute. 
        \item The paper should disclose whether the full research project required more compute than the experiments reported in the paper (e.g., preliminary or failed experiments that didn't make it into the paper). 
    \end{itemize}
    
\item {\bf Code of ethics}
    \item[] Question: Does the research conducted in the paper conform, in every respect, with the NeurIPS Code of Ethics \url{https://neurips.cc/public/EthicsGuidelines}?
    \item[] Answer: \answerYes{}
    \item[] Justification: The research uses publicly available MCP servers and open-source models. No human subjects or private data are involved.
    \item[] Guidelines:
    \begin{itemize}
        \item The answer \answerNA{} means that the authors have not reviewed the NeurIPS Code of Ethics.
        \item If the authors answer \answerNo, they should explain the special circumstances that require a deviation from the Code of Ethics.
        \item The authors should make sure to preserve anonymity (e.g., if there is a special consideration due to laws or regulations in their jurisdiction).
    \end{itemize}

\item {\bf Broader impacts}
    \item[] Question: Does the paper discuss both potential positive societal impacts and negative societal impacts of the work performed?
    \item[] Answer: \answerNA{}
    \item[] Justification: This work produces training data for tool-calling agents. The tools are publicly available APIs (weather, search, etc.) with no direct path to harmful applications beyond general LLM misuse risks.
    \item[] Guidelines:
    \begin{itemize}
        \item The answer \answerNA{} means that there is no societal impact of the work performed.
        \item If the authors answer \answerNA{} or \answerNo, they should explain why their work has no societal impact or why the paper does not address societal impact.
        \item Examples of negative societal impacts include potential malicious or unintended uses (e.g., disinformation, generating fake profiles, surveillance), fairness considerations (e.g., deployment of technologies that could make decisions that unfairly impact specific groups), privacy considerations, and security considerations.
        \item The conference expects that many papers will be foundational research and not tied to particular applications, let alone deployments. However, if there is a direct path to any negative applications, the authors should point it out. For example, it is legitimate to point out that an improvement in the quality of generative models could be used to generate Deepfakes for disinformation. On the other hand, it is not needed to point out that a generic algorithm for optimizing neural networks could enable people to train models that generate Deepfakes faster.
        \item The authors should consider possible harms that could arise when the technology is being used as intended and functioning correctly, harms that could arise when the technology is being used as intended but gives incorrect results, and harms following from (intentional or unintentional) misuse of the technology.
        \item If there are negative societal impacts, the authors could also discuss possible mitigation strategies (e.g., gated release of models, providing defenses in addition to attacks, mechanisms for monitoring misuse, mechanisms to monitor how a system learns from feedback over time, improving the efficiency and accessibility of ML).
    \end{itemize}
    
\item {\bf Safeguards}
    \item[] Question: Does the paper describe safeguards that have been put in place for responsible release of data or models that have a high risk for misuse (e.g., pre-trained language models, image generators, or scraped datasets)?
    \item[] Answer: \answerNA{}
    \item[] Justification: The released dataset contains tool-call trajectories from public APIs. The trained model is a 4B parameter model with no capabilities beyond its base checkpoint other than improved tool calling.
    \item[] Guidelines:
    \begin{itemize}
        \item The answer \answerNA{} means that the paper poses no such risks.
        \item Released models that have a high risk for misuse or dual-use should be released with necessary safeguards to allow for controlled use of the model, for example by requiring that users adhere to usage guidelines or restrictions to access the model or implementing safety filters. 
        \item Datasets that have been scraped from the Internet could pose safety risks. The authors should describe how they avoided releasing unsafe images.
        \item We recognize that providing effective safeguards is challenging, and many papers do not require this, but we encourage authors to take this into account and make a best faith effort.
    \end{itemize}

\item {\bf Licenses for existing assets}
    \item[] Question: Are the creators or original owners of assets (e.g., code, data, models), used in the paper, properly credited and are the license and terms of use explicitly mentioned and properly respected?
    \item[] Answer: \answerYes{}
    \item[] Justification: We cite the Qwen3 model family, the Smithery registry, and all benchmark datasets (Tau2-Bench, MCP-Atlas, MCPMark). MCP servers are accessed through their public APIs.
    \item[] Guidelines:
    \begin{itemize}
        \item The answer \answerNA{} means that the paper does not use existing assets.
        \item The authors should cite the original paper that produced the code package or dataset.
        \item The authors should state which version of the asset is used and, if possible, include a URL.
        \item The name of the license (e.g., CC-BY 4.0) should be included for each asset.
        \item For scraped data from a particular source (e.g., website), the copyright and terms of service of that source should be provided.
        \item If assets are released, the license, copyright information, and terms of use in the package should be provided. For popular datasets, \url{paperswithcode.com/datasets} has curated licenses for some datasets. Their licensing guide can help determine the license of a dataset.
        \item For existing datasets that are re-packaged, both the original license and the license of the derived asset (if it has changed) should be provided.
        \item If this information is not available online, the authors are encouraged to reach out to the asset's creators.
    \end{itemize}

\item {\bf New assets}
    \item[] Question: Are new assets introduced in the paper well documented and is the documentation provided alongside the assets?
    \item[] Answer: \answerYes{}
    \item[] Justification: The dataset, simulator code, and training scripts are released with documentation at \url{https://github.com/leoleoasd/firefly-neurips-2026/}, and the data and model at \url{http://firefly.hailab.io/}.
    \item[] Guidelines:
    \begin{itemize}
        \item The answer \answerNA{} means that the paper does not release new assets.
        \item Researchers should communicate the details of the dataset\slash code\slash model as part of their submissions via structured templates. This includes details about training, license, limitations, etc. 
        \item The paper should discuss whether and how consent was obtained from people whose asset is used.
        \item At submission time, remember to anonymize your assets (if applicable). You can either create an anonymized URL or include an anonymized zip file.
    \end{itemize}

\item {\bf Crowdsourcing and research with human subjects}
    \item[] Question: For crowdsourcing experiments and research with human subjects, does the paper include the full text of instructions given to participants and screenshots, if applicable, as well as details about compensation (if any)? 
    \item[] Answer: \answerNA{}
    \item[] Justification: This work does not involve crowdsourcing or human subjects.
    \item[] Guidelines:
    \begin{itemize}
        \item The answer \answerNA{} means that the paper does not involve crowdsourcing nor research with human subjects.
        \item Including this information in the supplemental material is fine, but if the main contribution of the paper involves human subjects, then as much detail as possible should be included in the main paper. 
        \item According to the NeurIPS Code of Ethics, workers involved in data collection, curation, or other labor should be paid at least the minimum wage in the country of the data collector. 
    \end{itemize}

\item {\bf Institutional review board (IRB) approvals or equivalent for research with human subjects}
    \item[] Question: Does the paper describe potential risks incurred by study participants, whether such risks were disclosed to the subjects, and whether Institutional Review Board (IRB) approvals (or an equivalent approval/review based on the requirements of your country or institution) were obtained?
    \item[] Answer: \answerNA{}
    \item[] Justification: This work does not involve human subjects.
    \item[] Guidelines:
    \begin{itemize}
        \item The answer \answerNA{} means that the paper does not involve crowdsourcing nor research with human subjects.
        \item Depending on the country in which research is conducted, IRB approval (or equivalent) may be required for any human subjects research. If you obtained IRB approval, you should clearly state this in the paper. 
        \item We recognize that the procedures for this may vary significantly between institutions and locations, and we expect authors to adhere to the NeurIPS Code of Ethics and the guidelines for their institution. 
        \item For initial submissions, do not include any information that would break anonymity (if applicable), such as the institution conducting the review.
    \end{itemize}

\item {\bf Declaration of LLM usage}
    \item[] Question: Does the paper describe the usage of LLMs if it is an important, original, or non-standard component of the core methods in this research? Note that if the LLM is used only for writing, editing, or formatting purposes and does \emph{not} impact the core methodology, scientific rigor, or originality of the research, declaration is not required.
    \item[] Answer: \answerYes{}
    \item[] Justification: LLMs are used as core components of the proposed pipeline, including tool-compatibility graph construction, graph-guided DAG exploration, back-chaining task synthesis, task validation, simulator response generation, and answer judging. These usages are described in Secs.~\ref{sec:tool_graph}, \ref{sec:exploration}, \ref{sec:backchain}, \ref{sec:validation}, and \ref{sec:simulator}. LLMs were also used for code writing.
    \begin{itemize}
        \item The answer \answerNA{} means that the core method development in this research does not involve LLMs as any important, original, or non-standard components.
        \item Please refer to our LLM policy in the NeurIPS handbook for what should or should not be described.
    \end{itemize}

\end{enumerate}

\end{document}

%% file: math_commands.tex
\usepackage{amsmath,amsfonts,bm}

\def\eqref#1{equation~\ref{#1}}

\def\1{\bm{1}}

\DeclareMathAlphabet{\mathsfit}{\encodingdefault}{\sfdefault}{m}{sl}
\SetMathAlphabet{\mathsfit}{bold}{\encodingdefault}{\sfdefault}{bx}{n}

%% file: main.bbl
\begin{thebibliography}{35}
\providecommand{\natexlab}[1]{#1}
\providecommand{\url}[1]{\texttt{#1}}
\expandafter\ifx\csname urlstyle\endcsname\relax
  \providecommand{\doi}[1]{doi: #1}\else
  \providecommand{\doi}{doi: \begingroup \urlstyle{rm}\Url}\fi

\bibitem[Bandi et~al.(2026)Bandi, Hertzberg, Boo, Polakam, Da, Hassaan, Sharma,
  Park, Hernandez, Rambado, Salazar, Cruz, Rane, Levin, Kenstler, and
  Liu]{bandiMCPAtlasLargeScaleBenchmark2026}
Chaithanya Bandi, Ben Hertzberg, Geobio Boo, Tejas Polakam, Jeff Da, Sami
  Hassaan, Manasi Sharma, Andrew Park, Ernesto Hernandez, Dan Rambado, Ivan
  Salazar, Rafael Cruz, Chetan Rane, Ben Levin, Brad Kenstler, and Bing Liu.
\newblock {{MCP-Atlas}}: {{A Large-Scale Benchmark}} for {{Tool-Use
  Competency}} with {{Real MCP Servers}}, May 2026.

\bibitem[Barres et~al.(2025)Barres, Dong, Ray, Si, and
  Narasimhan]{barrestau2BenchEvaluatingConversational2025}
Victor Barres, Honghua Dong, Soham Ray, Xujie Si, and Karthik Narasimhan.
\newblock {$\tau$}\textasciicircum 2-{{Bench}}: {{Evaluating Conversational
  Agents}} in a {{Dual-Control Environment}}, June 2025.

\bibitem[Chen et~al.(2026{\natexlab{a}})Chen, Nguyen, Zhang, Gullapalli, Wang,
  Lu, Wang, Huang, Yu, and Lai]{chenRetrySwitchAbstain2026}
Chaoran Chen, Vy~Nguyen, Ziji Zhang, Abhinav Gullapalli, Ziyi Wang, Yuxuan Lu,
  Dakuo Wang, Jing Huang, Zhou Yu, and Jin Lai.
\newblock Retry, {{Switch}}, or {{Abstain}}? {{Learning Strategy-Aware Tool-Use
  Policies}} via {{Controlled Error Injection}}, August 2026{\natexlab{a}}.

\bibitem[Chen et~al.(2024)Chen, Lu, Zhang, Yao, Dong, Xu, Li, Wang, Wang, and
  Sun]{chenStorySparkQAExpertAnnotatedQA2024a}
Jiaju Chen, Yuxuan Lu, Shao Zhang, Bingsheng Yao, Yuanzhe Dong, Ying Xu, Yunyao
  Li, Qianwen Wang, Dakuo Wang, and Yuling Sun.
\newblock {{StorySparkQA}}: {{Expert-Annotated QA Pairs}} with {{Real-World
  Knowledge}} for {{Children}}'s {{Story-Based Learning}}.
\newblock In Yaser {Al-Onaizan}, Mohit Bansal, and Yun-Nung Chen, editors,
  \emph{Proceedings of the 2024 {{Conference}} on {{Empirical Methods}} in
  {{Natural Language Processing}}}, pages 17351--17370, Miami, Florida, USA,
  November 2024. Association for Computational Linguistics.
\newblock \doi{10.18653/v1/2024.emnlp-main.961}.

\bibitem[Chen et~al.(2026{\natexlab{b}})Chen, Lu, Su, Chen, Xiao, Zhang, Wang,
  Li, Zhao, Wu, Li, Wang, and Yao]{chenHumansALMANACHuman2026}
Jiaju Chen, Yuxuan Lu, Jiayi Su, Chaoran Chen, Songlin Xiao, Zheng Zhang, Yun
  Wang, Yunyao Li, Jian Zhao, Tongshuang Wu, Toby Jia-Jun Li, Dakuo Wang, and
  Bingsheng Yao.
\newblock Humans' {{ALMANAC}}: {{A Human Collaboration Dataset}} of
  {{Action-Level Mental Model Annotations}} for {{Agent Collaboration}}, June
  2026{\natexlab{b}}.

\bibitem[Chen et~al.(2026{\natexlab{c}})Chen, Lu, Wang, Zeng, Huang, Gesi, Xu,
  Yao, and Wang]{chenMultiAgentasJudgeAligningLLMAgentBased2026}
Jiaju Chen, Yuxuan Lu, Xiaojie Wang, Huimin Zeng, Jing Huang, Jiri Gesi, Ying
  Xu, Bingsheng Yao, and Dakuo Wang.
\newblock Multi-{{Agent}}-as-{{Judge}}: {{Aligning LLM-Agent-Based Automated
  Evaluation}} with {{Multi-Dimensional Human Evaluation}}.
\newblock In Maria Liakata, Viviane~P. Moreira, Jiajun Zhang, and David
  Jurgens, editors, \emph{Proceedings of the 64th {{Annual Meeting}} of the
  {{Association}} for {{Computational Linguistics}} ({{Volume}} 1: {{Long
  Papers}})}, pages 17389--17416, San Diego, California, United States, July
  2026{\natexlab{c}}. Association for Computational Linguistics.
\newblock ISBN 979-8-89176-390-6.
\newblock \doi{10.18653/v1/2026.acl-long.790}.

\bibitem[Chen et~al.(2026{\natexlab{d}})Chen, Sun, Lu, Wang, Wang, and
  Yao]{chenCollabSimCSCWGroundedMethodology2026}
Jiaju Chen, Bo~Sun, Yuxuan Lu, Yun Wang, Dakuo Wang, and Bingsheng Yao.
\newblock {{CollabSim}}: {{A CSCW-Grounded Methodology}} for {{Investigating
  Collaborative Competence}} of {{LLM Agents}} through {{Controlled Multi-Agent
  Experiments}}, June 2026{\natexlab{d}}.

\bibitem[Chen et~al.(2021)Chen, Tworek, Jun, Yuan, Pinto, Kaplan, Edwards,
  Burda, Joseph, Brockman, Ray, Puri, Krueger, Petrov, Khlaaf, Sastry, Mishkin,
  Chan, Gray, Ryder, Pavlov, Power, Kaiser, Bavarian, Winter, Tillet, Such,
  Cummings, Plappert, Chantzis, Barnes, {Herbert-Voss}, Guss, Nichol, Paino,
  Tezak, Tang, Babuschkin, Balaji, Jain, Saunders, Hesse, Carr, Leike, Achiam,
  Misra, Morikawa, Radford, Knight, Brundage, Murati, Mayer, Welinder, McGrew,
  Amodei, McCandlish, Sutskever, and Zaremba]{chenEvaluatingLargeLanguage2021}
Mark Chen, Jerry Tworek, Heewoo Jun, Qiming Yuan, Henrique Ponde de~Oliveira
  Pinto, Jared Kaplan, Harri Edwards, Yuri Burda, Nicholas Joseph, Greg
  Brockman, Alex Ray, Raul Puri, Gretchen Krueger, Michael Petrov, Heidy
  Khlaaf, Girish Sastry, Pamela Mishkin, Brooke Chan, Scott Gray, Nick Ryder,
  Mikhail Pavlov, Alethea Power, Lukasz Kaiser, Mohammad Bavarian, Clemens
  Winter, Philippe Tillet, Felipe~Petroski Such, Dave Cummings, Matthias
  Plappert, Fotios Chantzis, Elizabeth Barnes, Ariel {Herbert-Voss},
  William~Hebgen Guss, Alex Nichol, Alex Paino, Nikolas Tezak, Jie Tang, Igor
  Babuschkin, Suchir Balaji, Shantanu Jain, William Saunders, Christopher
  Hesse, Andrew~N. Carr, Jan Leike, Josh Achiam, Vedant Misra, Evan Morikawa,
  Alec Radford, Matthew Knight, Miles Brundage, Mira Murati, Katie Mayer, Peter
  Welinder, Bob McGrew, Dario Amodei, Sam McCandlish, Ilya Sutskever, and
  Wojciech Zaremba.
\newblock Evaluating {{Large Language Models Trained}} on {{Code}}, July 2021.

\bibitem[Chen et~al.(2025)Chen, Zhao, Zhang, Liu, Qi, Wu, Kalluri, Cao, Xiong,
  Tong, Yao, Li, Zhu, Li, Song, Li, Weston, and
  Huynh]{chenScalingAgentLearning2025}
Zhaorun Chen, Zhuokai Zhao, Kai Zhang, Bo~Liu, Qi~Qi, Yifan Wu, Tarun Kalluri,
  Sara Cao, Yuanhao Xiong, Haibo Tong, Huaxiu Yao, Hengduo Li, Jiacheng Zhu,
  Xian Li, Dawn Song, Bo~Li, Jason Weston, and Dat Huynh.
\newblock Scaling {{Agent Learning}} via {{Experience Synthesis}}, November
  2025.

\bibitem[Jiang et~al.(2025)Jiang, Xu, Lin, Xiao, Wang, Sun, and
  Han]{jiangS3YouDont2025}
Pengcheng Jiang, Xueqiang Xu, Jiacheng Lin, Jinfeng Xiao, Zifeng Wang, Jimeng
  Sun, and Jiawei Han.
\newblock S3: {{You Don}}'t {{Need That Much Data}} to {{Train}} a {{Search
  Agent}} via {{RL}}, November 2025.

\bibitem[Li et~al.(2023)Li, Zhao, Yu, Song, Li, Yu, Li, Huang, and
  Li]{liAPIBankComprehensiveBenchmark2023}
Minghao Li, Yingxiu Zhao, Bowen Yu, Feifan Song, Hangyu Li, Haiyang Yu, Zhoujun
  Li, Fei Huang, and Yongbin Li.
\newblock {{API-Bank}}: {{A Comprehensive Benchmark}} for {{Tool-Augmented
  LLMs}}, October 2023.

\bibitem[Lu et~al.(2025{\natexlab{a}})Lu, Chen, Zhang, Shen, Wang, Wang, van
  Rechem, Fu, and Wei]{luMachineLearningSynthetic2025}
Yingzhou Lu, Lulu Chen, Yuanyuan Zhang, Minjie Shen, Huazheng Wang, Xiao Wang,
  Capucine van Rechem, Tianfan Fu, and Wenqi Wei.
\newblock Machine {{Learning}} for {{Synthetic Data Generation}}: {{A Review}},
  April 2025{\natexlab{a}}.

\bibitem[Lu et~al.(2025{\natexlab{b}})Lu, Huang, Liu, Gesi, Han, Fu, Zheng, and
  Wang]{luWEBSERVBrowserServerEnvironment2025a}
Yuxuan Lu, Jing Huang, Hui Liu, Jiri Gesi, Yan Han, Shihan Fu, Tianqi Zheng,
  and Dakuo Wang.
\newblock {{WEBSERV}}: {{A Browser-Server Environment}} for {{Efficient
  Training}} of {{Reinforcement Learning}}-based {{Web Agents}} at {{Scale}}.
\newblock In \emph{First {{Workshop}} on {{Multi-Turn Interactions}} in {{Large
  Language Models}}}, 2025{\natexlab{b}}.
\newblock URL \url{https://openreview.net/forum?id=YHcHQY8TIY}.

\bibitem[Lu et~al.(2026)Lu, Yao, Zhang, Sang, Wang, Gu, Zhang, Lu, Li, and
  Wang]{luLLMsNowFineTuning2026a}
Yuxuan Lu, Bingsheng Yao, Shao Zhang, Yisi Sang, Yun Wang, Hansu Gu, Peng
  Zhang, Tun Lu, Toby Jia-Jun Li, and Dakuo Wang.
\newblock {{LLMs}} for {{Now}}, {{Fine-Tuning}} for {{Later}}: {{An Ensemble
  Approach}} to {{Data Drift}} in {{Domain-Specific Tasks}}.
\newblock In Santosh T.Y.S.S., Juan~Diego Rodriguez, and Ona {de Gibert},
  editors, \emph{Proceedings of the 64th {{Annual Meeting}} of the
  {{Association}} for {{Computational Linguistics}} ({{Volume}} 4: {{Student
  Research Workshop}})}, pages 861--876, San Diego, California, United States,
  July 2026. Association for Computational Linguistics.
\newblock ISBN 979-8-89176-393-7.
\newblock \doi{10.18653/v1/2026.acl-srw.77}.

\bibitem[Luo et~al.(2025)Luo, Sun, Xu, Zhao, Lou, Tao, Geng, Lin, Chen, Tang,
  and Zhang]{luoWizardMathEmpoweringMathematical2025}
Haipeng Luo, Qingfeng Sun, Can Xu, Pu~Zhao, Jianguang Lou, Chongyang Tao, Xiubo
  Geng, Qingwei Lin, Shifeng Chen, Yansong Tang, and Dongmei Zhang.
\newblock {{WizardMath}}: {{Empowering Mathematical Reasoning}} for {{Large
  Language Models}} via {{Reinforced Evol-Instruct}}, June 2025.

\bibitem[Prabhakar et~al.(2025)Prabhakar, Liu, Zhu, Zhang, Awalgaonkar, Wang,
  Liu, Chen, Hoang, Niebles, Heinecke, Yao, Wang, Savarese, and
  Xiong]{prabhakarAPIGenMTAgenticPipeline2025a}
Akshara Prabhakar, Zuxin Liu, Ming Zhu, Jianguo Zhang, Tulika Awalgaonkar,
  Shiyu Wang, Zhiwei Liu, Haolin Chen, Thai Hoang, Juan~Carlos Niebles, Shelby
  Heinecke, Weiran Yao, Huan Wang, Silvio Savarese, and Caiming Xiong.
\newblock {{APIGen-MT}}: {{Agentic Pipeline}} for {{Multi-Turn Data
  Generation}} via {{Simulated Agent-Human Interplay}}, July 2025.

\bibitem[Qi et~al.(2025)Qi, Zhao, Guo, Ma, Chen, Han, Zhang, Lin, Xin, Huang,
  et~al.]{qi2025bear}
Yu~Qi, Haibo Zhao, Ziyu Guo, Siyuan Ma, Ziyan Chen, Yaokun Han, Renrui Zhang,
  Zitiantao Lin, Shiji Xin, Yijian Huang, et~al.
\newblock Bear: Benchmarking and enhancing multimodal language models for
  atomic embodied capabilities.
\newblock \emph{arXiv e-prints}, pages arXiv--2510, 2025.

\bibitem[Rozi{\`e}re et~al.(2023)Rozi{\`e}re, Gehring, Gloeckle, Sootla, Gat,
  Tan, Adi, Liu, Remez, Rapin, Kozhevnikov, Evtimov, Bitton, Bhatt, Ferrer,
  Grattafiori, Xiong, D{\'e}fossez, Copet, Azhar, Touvron, Martin, Usunier,
  Scialom, and Synnaeve]{roziereCodeLlamaOpen2023}
Baptiste Rozi{\`e}re, Jonas Gehring, Fabian Gloeckle, Sten Sootla, Itai Gat,
  Xiaoqing~Ellen Tan, Yossi Adi, Jingyu Liu, Tal Remez, J{\'e}r{\'e}my Rapin,
  Artyom Kozhevnikov, Ivan Evtimov, Joanna Bitton, Manish Bhatt,
  Cristian~Canton Ferrer, Aaron Grattafiori, Wenhan Xiong, Alexandre
  D{\'e}fossez, Jade Copet, Faisal Azhar, Hugo Touvron, Louis Martin, Nicolas
  Usunier, Thomas Scialom, and Gabriel Synnaeve.
\newblock Code {{Llama}}: {{Open Foundation Models}} for {{Code}}, August 2023.

\bibitem[Schick et~al.(2023)Schick, {Dwivedi-Yu}, Dessi, Raileanu, Lomeli,
  Hambro, Zettlemoyer, Cancedda, and
  Scialom]{schickToolformerLanguageModels2023}
Timo Schick, Jane {Dwivedi-Yu}, Roberto Dessi, Roberta Raileanu, Maria Lomeli,
  Eric Hambro, Luke Zettlemoyer, Nicola Cancedda, and Thomas Scialom.
\newblock Toolformer: {{Language Models Can Teach Themselves}} to {{Use
  Tools}}.
\newblock \emph{Advances in Neural Information Processing Systems},
  36:\penalty0 68539--68551, December 2023.
\newblock URL
  \url{https://proceedings.neurips.cc/paper_files/paper/2023/hash/d842425e4bf79ba039352da0f658a906-Abstract-Conference.html}.

\bibitem[Shao et~al.(2024)Shao, Wang, Zhu, Xu, Song, Bi, Zhang, Zhang, Li, Wu,
  and Guo]{shaoDeepSeekMathPushingLimits2024}
Zhihong Shao, Peiyi Wang, Qihao Zhu, Runxin Xu, Junxiao Song, Xiao Bi, Haowei
  Zhang, Mingchuan Zhang, Y.~K. Li, Y.~Wu, and Daya Guo.
\newblock {{DeepSeekMath}}: {{Pushing}} the {{Limits}} of {{Mathematical
  Reasoning}} in {{Open Language Models}}, April 2024.

\bibitem[Tang et~al.(2023)Tang, Deng, Lin, Han, Liang, Cao, and
  Sun]{tangToolAlpacaGeneralizedTool2023}
Qiaoyu Tang, Ziliang Deng, Hongyu Lin, Xianpei Han, Qiao Liang, Boxi Cao, and
  Le~Sun.
\newblock {{ToolAlpaca}}: {{Generalized Tool Learning}} for {{Language Models}}
  with 3000 {{Simulated Cases}}, September 2023.

\bibitem[Wang et~al.(2026{\natexlab{a}})Wang, Niu, Xu, Chen, Du, Du, Pang,
  Huang, Wang, Yan, and Chen]{wangMCPFlowFacilitatingLLM2026}
Wenhao Wang, Peizhi Niu, Zhao Xu, Zhaoyu Chen, Jian Du, Yaxin Du, Xianghe Pang,
  Keduan Huang, Yanfeng Wang, Qiang Yan, and Siheng Chen.
\newblock {{MCP-Flow}}: {{Facilitating LLM Agents}} to {{Master Real-World}},
  {{Diverse}} and {{Scaling MCP Tools}}, April 2026{\natexlab{a}}.

\bibitem[Wang et~al.(2023)Wang, Wei, Schuurmans, Le, Chi, Narang, Chowdhery,
  and Zhou]{wangSelfConsistencyImprovesChain2023}
Xuezhi Wang, Jason Wei, Dale Schuurmans, Quoc Le, Ed~Chi, Sharan Narang,
  Aakanksha Chowdhery, and Denny Zhou.
\newblock Self-{{Consistency Improves Chain}} of {{Thought Reasoning}} in
  {{Language Models}}, March 2023.

\bibitem[Wang et~al.(2025)Wang, Lu, Zhang, Huang, and
  Wang]{wangCustomerR1PersonalizedSimulation2025a}
Ziyi Wang, Yuxuan Lu, Yimeng Zhang, Jing Huang, and Dakuo Wang.
\newblock Customer-{{R1}}: {{Personalized Simulation}} of {{Human Behaviors}}
  via {{RL}}-based {{LLM Agent}} in {{Online Shopping}}, October 2025.

\bibitem[Wang et~al.(2026{\natexlab{b}})Wang, Lu, Zhang, Chen, Dong, Huang,
  Gesi, Tang, Luo, and Liu]{wangTrajectory2TaskTrainingRobust2026}
Ziyi Wang, Yuxuan Lu, Yimeng Zhang, Pei Chen, Ziwei Dong, Jing Huang, Jiri
  Gesi, Xianfeng Tang, Chen Luo, and Qun Liu.
\newblock Trajectory2task: {{Training}} robust tool-calling agents with
  synthesized yet verifiable data for complex user intents.
\newblock In \emph{Proceedings of the 64th {{Annual Meeting}} of the
  {{Association}} for {{Computational Linguistics}} ({{Volume}} 1: {{Long
  Papers}})}, pages 44021--44044, 2026{\natexlab{b}}.
\newblock URL \url{https://aclanthology.org/2026.acl-long.2037/}.

\bibitem[Wang et~al.(2026{\natexlab{c}})Wang, Lu, Zhang, Liu, Luo, Gesi, Lu,
  Sang, Li, Huang, and Wang]{wangSENTINELFailureDrivenReinforcement2026}
Ziyi Wang, Yuxuan Lu, Yimeng Zhang, Qun Liu, Chen Luo, Jiri Gesi, Hanqing Lu,
  Yisi Sang, Manling Li, Jing Huang, and Dakuo Wang.
\newblock {{SENTINEL}}: {{Failure-Driven Reinforcement Learning}} for
  {{Training Tool-Using Language Model Agents}}, June 2026{\natexlab{c}}.

\bibitem[Wei et~al.(2023)Wei, Wang, Schuurmans, Bosma, Ichter, Xia, Chi, Le,
  and Zhou]{weiChainThoughtPromptingElicits2023}
Jason Wei, Xuezhi Wang, Dale Schuurmans, Maarten Bosma, Brian Ichter, Fei Xia,
  Ed~Chi, Quoc Le, and Denny Zhou.
\newblock Chain-of-{{Thought Prompting Elicits Reasoning}} in {{Large Language
  Models}}, January 2023.

\bibitem[Wu et~al.(2025)Wu, Liu, Zhang, Chen, Meng, Du, Zhao, Zhang, Ye, Wang,
  Wang, Ni, Yang, Xu, and Shieh]{wuMCPMarkBenchmarkStressTesting2025}
Zijian Wu, Xiangyan Liu, Xinyuan Zhang, Lingjun Chen, Fanqing Meng, Lingxiao
  Du, Yiran Zhao, Fanshi Zhang, Yaoqi Ye, Jiawei Wang, Zirui Wang, Jinjie Ni,
  Yufan Yang, Arvin Xu, and Michael~Qizhe Shieh.
\newblock {{MCPMark}}: {{A Benchmark}} for {{Stress-Testing Realistic}} and
  {{Comprehensive MCP Use}}, September 2025.

\bibitem[Xu et~al.(2025{\natexlab{a}})Xu, Sun, Zheng, Geng, Zhao, Feng, Tao,
  Lin, and Jiang]{xuWizardLMEmpoweringLarge2025}
Can Xu, Qingfeng Sun, Kai Zheng, Xiubo Geng, Pu~Zhao, Jiazhan Feng, Chongyang
  Tao, Qingwei Lin, and Daxin Jiang.
\newblock {{WizardLM}}: {{Empowering}} large pre-trained language models to
  follow complex instructions, May 2025{\natexlab{a}}.

\bibitem[Xu et~al.(2024)Xu, Li, Tao, Shen, Cheng, Li, Xu, Tao, and
  Zhou]{xuSurveyKnowledgeDistillation2024}
Xiaohan Xu, Ming Li, Chongyang Tao, Tao Shen, Reynold Cheng, Jinyang Li, Can
  Xu, Dacheng Tao, and Tianyi Zhou.
\newblock A {{Survey}} on {{Knowledge Distillation}} of {{Large Language
  Models}}, October 2024.

\bibitem[Xu et~al.(2025{\natexlab{b}})Xu, Soria, Tan, Roy, Agrawal, Poovendran,
  and Panda]{xuTOUCANSynthesizing15M2025}
Zhangchen Xu, Adriana~Meza Soria, Shawn Tan, Anurag Roy, Ashish~Sunil Agrawal,
  Radha Poovendran, and Rameswar Panda.
\newblock {{TOUCAN}}: {{Synthesizing}} 1.{{5M Tool-Agentic Data}} from
  {{Real-World MCP Environments}}, October 2025{\natexlab{b}}.

\bibitem[Yang et~al.(2025)Yang, Li, Yang, Zhang, Hui, Zheng, Yu, Gao, Huang,
  Lv, Zheng, Liu, Zhou, Huang, Hu, Ge, Wei, Lin, Tang, Yang, Tu, Zhang, Yang,
  Yang, Zhou, Zhou, Lin, Dang, Bao, Yang, Yu, Deng, Li, Xue, Li, Zhang, Wang,
  Zhu, Men, Gao, Liu, Luo, Li, Tang, Yin, Ren, Wang, Zhang, Ren, Fan, Su,
  Zhang, Zhang, Wan, Liu, Wang, Cui, Zhang, Zhou, and
  Qiu]{yangQwen3TechnicalReport2025}
An~Yang, Anfeng Li, Baosong Yang, Beichen Zhang, Binyuan Hui, Bo~Zheng, Bowen
  Yu, Chang Gao, Chengen Huang, Chenxu Lv, Chujie Zheng, Dayiheng Liu, Fan
  Zhou, Fei Huang, Feng Hu, Hao Ge, Haoran Wei, Huan Lin, Jialong Tang, Jian
  Yang, Jianhong Tu, Jianwei Zhang, Jianxin Yang, Jiaxi Yang, Jing Zhou,
  Jingren Zhou, Junyang Lin, Kai Dang, Keqin Bao, Kexin Yang, Le~Yu, Lianghao
  Deng, Mei Li, Mingfeng Xue, Mingze Li, Pei Zhang, Peng Wang, Qin Zhu, Rui
  Men, Ruize Gao, Shixuan Liu, Shuang Luo, Tianhao Li, Tianyi Tang, Wenbiao
  Yin, Xingzhang Ren, Xinyu Wang, Xinyu Zhang, Xuancheng Ren, Yang Fan, Yang
  Su, Yichang Zhang, Yinger Zhang, Yu~Wan, Yuqiong Liu, Zekun Wang, Zeyu Cui,
  Zhenru Zhang, Zhipeng Zhou, and Zihan Qiu.
\newblock Qwen3 {{Technical Report}}, May 2025.

\bibitem[Yao et~al.(2024)Yao, Shinn, Razavi, and
  Narasimhan]{yaotaubenchBenchmarkToolAgentUser2024}
Shunyu Yao, Noah Shinn, Pedram Razavi, and Karthik Narasimhan.
\newblock {$\tau$}-bench: {{A Benchmark}} for {{Tool-Agent-User Interaction}}
  in {{Real-World Domains}}, June 2024.

\bibitem[Yu et~al.(2025)Yu, Zhang, Zhu, Yuan, Zuo, Yue, Dai, Fan, Liu, Liu,
  Liu, Lin, Lin, Ma, Sheng, Tong, Zhang, Zhang, Zhang, Zhu, Zhu, Chen, Chen,
  Wang, Yu, Song, Wei, Zhou, Liu, Ma, Zhang, Yan, Qiao, Wu, and
  Wang]{yuDAPOOpenSourceLLM2025}
Qiying Yu, Zheng Zhang, Ruofei Zhu, Yufeng Yuan, Xiaochen Zuo, Yu~Yue, Weinan
  Dai, Tiantian Fan, Gaohong Liu, Lingjun Liu, Xin Liu, Haibin Lin, Zhiqi Lin,
  Bole Ma, Guangming Sheng, Yuxuan Tong, Chi Zhang, Mofan Zhang, Wang Zhang,
  Hang Zhu, Jinhua Zhu, Jiaze Chen, Jiangjie Chen, Chengyi Wang, Hongli Yu,
  Yuxuan Song, Xiangpeng Wei, Hao Zhou, Jingjing Liu, Wei-Ying Ma, Ya-Qin
  Zhang, Lin Yan, Mu~Qiao, Yonghui Wu, and Mingxuan Wang.
\newblock {{DAPO}}: {{An Open-Source LLM Reinforcement Learning System}} at
  {{Scale}}, May 2025.

\bibitem[Zhang et~al.(2026)Zhang, Wang, Gesi, Wang, Lu, Lin, Zhan, Gao, Jiao,
  and Liu]{zhangShopr1RewardingLlms2026}
Yimeng Zhang, Tian Wang, Jiri Gesi, Ziyi Wang, Yuxuan Lu, Jiacheng Lin, Simon
  Zhan, Vianne Gao, Ruochen Jiao, and Junze Liu.
\newblock Shop-r1: {{Rewarding}} llms to simulate human behavior in online
  shopping via reinforcement learning.
\newblock In \emph{International {{Conference}} on {{Learning
  Representations}}}, volume 2026, pages 8406--8423, 2026.
\newblock URL
  \url{https://proceedings.iclr.cc/paper_files/paper/2026/hash/0e705ac30e573d1526f81a0fd071a151-Abstract-Conference.html}.

\end{thebibliography}
